\documentclass[article,usenatbib]{mn2e}
\usepackage[english,activeacute]{babel}
\usepackage{setspace}
\usepackage{graphics}
\usepackage{graphicx}
\usepackage{amssymb}
\usepackage{hyperref}
\usepackage{longtable}
\usepackage{amsmath}
\usepackage{indentfirst}
\usepackage{verbatim}
\usepackage{array}
\usepackage{amssymb}
\usepackage{subfigure}
\usepackage{fancyhdr}
\usepackage{amsmath}
\usepackage{verbatim}

\begin{document}
\title[Dark Matter annihilation energy output and its effects on the high-z IGM]{Dark Matter annihilation energy output and its effects on the high-z IGM}
\author[Ignacio J. Araya and Nelson D. Padilla]{Ignacio J. Araya$^{1}$ and Nelson D. Padilla$^{2}$\\
$^{1}$Department of Physics and Astronomy, University of Southern California, 920 Bloom Walk, Los Angeles, California, U.S.\\
$^{2}$Departamento de Astronom\'ia y Astrof\'isica, Pontificia Universidad Cat\'olica de Chile, Av. Vicu\~na Mackenna 4860, Stgo., Chile}
\maketitle

\begin{abstract}
We study the case of DM self annihilation, in order to { assess} its importance as an energy injection mechanism, to the IGM in general, and to the medium within particular DM haloes. { We consider thermal relic WIMP particles with masses of $10GeV$ and $1TeV$ and} we analyse in detail the clustering properties of { DM in a $\Lambda$CDM cosmology}, on all hierarchy levels, from haloes { and their mass function}, to subhaloes and { the} DM density profiles within them, { considering adiabatic contraction by the presence of a SMBH}. { We then compute the corresponding energy output, concluding that DM annihilation does not constitute an important feedback mechanism.} { We also calculate the effects that DM annihilation has on the IGM temperature and ionization fraction, and we find that assuming maximal energy absorption, at $z\sim 10$, for the case of a $1TeV$ WIMP, the ionization fraction could be raised to $6 \times 10^{-4}$ and the temperature to $10K$, and in the case of a $10GeV$ WIMP, the IGM temperature could be raised to $200K$ and the ionization fraction to $8 \times 10^{-3} $.} We conclude that DM annihilations cannot be regarded as an alternative reionization scenario. Regarding the detectability of the WIMP through the modifications to the 21 cm differential brightness temperature signal ($\delta T_{b}$), we conclude that { a thermal relic WIMP with mass of $1TeV$ is not likely to be detected from the global signal alone, except perhaps at the $1-3mK$ level in the frequency range $30MHz < \nu < 35MHz$ corresponding to $40 < z < 46$. However, a $10GeV$ mass WIMP may be detectable at the $1-3mK$ level in the frequency range $55MHz < \nu < 119MHz$ corresponding to $11 < z < 25$, and at the $1-10mK$ level in the frequency range $30MHz < \nu < 40MHz$ corresponding to $35 < z < 46$.}
\end{abstract}


\section{Introduction}

The Dark Matter (DM) is a component of the total density of the universe which can be measured only through gravitational interactions and (likely) weak interactions. It behaves as dust-like matter in the sense that it is collisionless and that its energy density decreases as the third power of the scale factor.
There are many compelling evidences of its existence, although all of them are indirect. These are mainly dynamical evidences, which rely on the gravitational influence of DM. Some examples of these evidences are the rotation curves of late type galaxies \citep{Rubin}, galaxy cluster dynamics \citep{Zwicky} and strong lensing \citep{Paczynski}, among others. The effects of DM are very important not only for determining the dynamics of galaxies or galaxy clusters, but also for the evolution of the universe as a whole, and for the galaxy formation processes and structure formation scenarios.

For example, assuming that the DM consists of particles that were originally in thermal equilibrium with the radiation in the early universe, depending on the mass of the particles, the structure formation scenario will be top down or hierarchical. In the top down structure formation scenario, the first structures to collapse are large (on the mass scale that corresponds to present-day superclusters), because all overdensities at smaller scales are erased by free streaming. This is the case for a hot dark matter (HDM) model, in which the DM particles were relativistic at the time of decoupling. In the hierarchical structure formation scenario, small (and also large) structures start to collapse after decoupling, because DM particles were non-relativistic at the time of decoupling, and so the free streaming is negligible. This is the case for any cold dark matter (CDM) model \citep{Blumenthal}, including the currently favoured $\Lambda CDM$ model (see e.g. the Planck results, Ade et al. 2013), which we study in this work.

The postulated CDM particle has been named a WIMP (or Weakly Interacting Massive Particle), because it weakly couples to normal matter (for introductions to the topic of DM astrophysics, see \cite{DAmico} and \cite{Raffelt}). In the standard scenario, this particle is considered to be a thermal relic. This means that its creation and annihilation processes from the early universe radiation field were in thermal equilibrium with each other, and thus, the present day DM number density is the relic abundance after the freeze-out of this processes, with annihilation being favoured after the universe had cooled enough \citep{EarlyUniverse}. This fact alone allows for the computation of the interaction cross section for the creation-annihilation process given only the present day abundance of DM (its density parameter $\Omega_{m,0}$).

An important feature of a self-annihilating thermal relic WIMP is that the annihilation products are standard model particles, and thus, they may inject energy to the IGM or to the medium where a particular DM halo is embedded. Globally, the annihilations of WIMPs are unlikely to occur because their abundance has long ago frozen. However, there may be regions where encounters among WIMPs are more probable due to an enhanced density. These regions are precisely the regions where structures form. Now, it is a direct consequence of the $\Lambda CDM$ model that DM is clumpy, due to the hierarchical growth of structure. Thus, it is precisely in the inner regions of DM haloes and substructures that the annihilations are more probable. The energy output due to annihilations will be proportional to the square of the density, thus, these local density enhancements are of great importance \citep{NatarajanCrotonBertone}.

The first hierarchy in density enhancements are the DM haloes that do not belong to any gravitationally bound structure at a particular redshift. { Their average clustering can be characterized by the halo mass function, which can be computed by Press-Schecter theory (e.g., \cite{PressSchechter}; \cite{MoWhite}; \cite{BondEtAl}; \cite{SMT}; \cite{LaceyCole} among others)}.

The second hierarchy in density enhancements are the substructures embedded in a particular halo. These subhaloes will retain their identities { even after coalescing}. Thus, inside each DM halo, it is expected that the full hierarchy of DM haloes is nested within, until the lowest mass haloes that formed. The number density of these subhaloes can be measured in Milky-Way sized halo simulations, and by approximating their abundance by a power-law behaviour (Via Lactea II simulation \citep{Diemand}, Acquarius project simulation \citep{Springel}).

Apart from the number density of haloes and subhaloes, it is important to quantify the form of the density distribution within a particular structure, in order to compute the annihilation power output. { An example of such a distribution is given by the NFW profile \citep{NFW}}. Also, this DM density profile may be further enhanced by the gravitational effect of baryons on DM through adiabatic contraction (see, for example \cite{SellwoodAndMcGaugh}; and \cite{NatarajanCrotonBertone}). Thus, the total resulting DM distribution in the universe can be computed at a particular redshift.

Knowing the global DM distribution, one can compute the resulting energy output per baryon, which can be further compared with what can be obtained assuming a perfectly homogeneous DM distribution (see \cite{MFP06}; and \cite{CLS}). One can also compute the energy generated per unit time by a particular halo of a given mass and redshift, and one can calculate what is the energy received by a particular halo due to the rest of the universe.

This additional energy injection may have interesting effects on the global IGM, as well as on the environment within particular DM haloes. For example, several authors (for example \cite{CLS}; \cite{MFP06}; \cite{RMF07}; \cite{VFMR07}; \cite{Pierpaoli}; \cite{NatarajanAndSchwarz}; {  \cite{Furlanetto}; \cite{Chuzhoy} among others)} have studied the effects of DM annihilation (and also decays, in different DM models) on the IGM temperature and on the global ionization fraction. This extra heating may be detectable through the global 21 cm differential brightness temperature, measurable in the MHz frequency range. This differential brightness temperature arises as a result of the coupling of the neutral hydrogen spin temperature (that accounts for the hyperfine transition) to the IGM kinetic temperature, thus causing a difference between the spin temperature and the CMB temperature.
 Also, the energy input may contribute as an extra heating mechanism within haloes that can add up to the standard SNe and AGN feedback mechanisms (\cite{NatarajanCrotonBertone}; \cite{Ascasibar}), and may change the temperature of the halo environment.

Regarding the observational evidence for the energy injection due to DM annihilations, no definite conclusion has been drawn yet. However, different observations of anomalous enhancements in detected signals from high-energy phenomena in the galactic halo and in the cosmic background suggest its presence.  These signals correspond mainly to gamma ray detections (Fermi \citep{FERMI}, H.E.S.S \citep{HESS} and EGRET \citep{EGRET} experiments) and to measurements of cosmic ray positron fractions (PAMELA \citep{PAMELA} experiment). For example, \cite{Cline} and \cite{FinkbeinerEtAl} analyse the data sets by PAMELA, Fermi and H.E.S.S., and conclude that the observed signals could be accounted for by a $\sim 1TeV$ WIMP. They also find an enhanced interaction cross-section (Sommerfeld enhancement) for the annihilation process ($\sim 10^{2}$ times the thermal relic WIMP value). A similar result was also obtained by \cite{Elsasser} with respect to the EGRET gamma ray data { (although there are other interpretations of it, for example in terms of $\sim 60GeV$ massive neutrinos by \cite{Belotsky})}. However, there is some debate regarding the compatibility of the presence of Sommerfeld enhancement with the non-detection of gamma ray flux from known DM structures (like galaxy clusters). In this line, the study by \cite{Pinzke} determines that the attempts of explaining the Fermi, PAMELA and H.E.S.S. combined data with $\sim 1TeV$ WIMPs and cross sections enhanced by factors of $10^{2}-10^{3}$, are incompatible with the EGRET upper limits on the gamma ray emission from the Virgo galaxy cluster. Regarding the Fermi gamma ray excess, \cite{HooperAndGoodenough} argue that it can be well accounted for by non-exotic processes (decaying pions, inverse Compton, point sources, etc...), except within $1.25^{\circ}$ ($\approx 175 pc$) from the galactic center, where the gamma ray excess is consistent with the prediction of a cusped ($\rho \propto r^{-1.34}$) DM density profile, and annihilating WIMPs with a thermal relic cross-section (no Sommerfeld enhancement) and masses in the range $m_{\chi} = 7.3GeV - 9.2GeV$. Also, \cite{HooperFinkbeinerDobler} found that the excess microwave radiation within $20^{\circ}$ from the galactic center, found by the WMAP experiment and called the WMAP Haze, can be well explained by a cusped DM profile with $\rho \propto r^{-1.2}$ in the inner kiloparsecs, and a WIMP with mass in the $100GeV$ to multi-$TeV$ range and thermal relic cross section. However, according to the recent null results in DM direct detection obtained by the LUX collaboration \citep{LUX}, a WIMP with masses between $\sim 10GeV$ an $\sim 100GeV$ is increasingly disfavoured. Also, for lower mass light dark matter (LDM) particle candidates (with masses in the $1MeV-100MeV$ range), possible constraints on the mass and cross section coming from the CMB have been studied (see section 5 of \cite{MFP06}; and \cite{Zhang}). However, the effect on the CMB by massive WIMPs, like the ones studied in this work, would be completely negligible. It can be seen that no clear consensus exists neither on the significance of the results as positive indirect detections of DM, nor on the importance of these detections for discriminating among different WIMP scenarios. What can be said nonetheless, is that evidence of possible new exotic physics is accumulating, and that WIMP DM annihilation represents a well motivated option.

The main goal of this work is to compute this additional energy injection, both locally within haloes and globally, and to { assess} the importance of this injection on the different processes mentioned above. { All the computations will be done considering a cosmology consistent with the Planck 2013 results \citep{Planck}. In particular, we will use $h = 0.6711$, $\Omega_{m,0} = 0.3174$, $\Omega_{\Lambda,0} = 0.6825$, $f_{bary} \equiv \frac{\Omega_{b,0}}{\Omega_{m,0}} = 0.17$, $n_{s} = 0.9624$ and $\sigma_{8} = 0.8344$.} The paper is organized as follows. In Section 2, the thermal relic WIMP, and the two above mentioned well motivated particular DM models will be explained in greater detail. In Section 3, the statistics of the DM clustering on all hierarchies will be developed and the NFW universal density profile, as well as the adiabatic contraction mechanism will be explained. In Section 4, the DM annihilation energy output, both globally (per baryon) and locally within a particular DM halo will be computed. Also, the clumpiness factor, defined as the ratio of output between the smoothed and fully clumped cases, will be calculated. In Section 5, { the intensity of the radiation field will be computed, and the power received from it by a halo will be obtained.} In Section 6, the effects of the annihilation energy injection on the global IGM will be followed in detail, computing the IGM temperature, the ionization fraction and the resulting 21 cm differential brightness temperature in this scenario. The general conclusions will be presented in Section 7.



\section{WIMP models and the thermal relic}

The WIMP is a theorized particle that allows to account for the missing mass in the universe. The evidence of this missing mass is mainly gravitational. This missing mass cannot come however from normal baryonic matter, because current Big-Bang nucleosynthesis constraints, like for example the primordial Helium abundance (Section 10.4 of \cite{Ryden}), set stringent limits on the normal matter content.

For a certain particle, motivated by some beyond-standard physical scenario, to be a viable WIMP candidate, it must have some basic properties. For example, it has to be charge-neutral and it can interact only weakly with ordinary matter. Also, it has to be massive in order to behave as Cold Dark Matter. A light DM particle will behave as Warm or Hot dark matter depending on if it was relativistic or not at the time of decoupling and also depending on its free streaming length \citep{Angulo}, and the resulting structure formation scenario or minimum halo mass will be different to the CDM case. Also, a WIMP must be stable, or have a decay time longer than (or at least comparable to) the present Hubble time. 

In some well motivated physical scenarios, the WIMP is a thermal relic. This means, as was mentioned before, that its creation and annihilation reaction were in thermal equilibrium with the radiation background in the early universe. Then, as the universe cooled down, the annihilation reaction was increasingly favoured over the creation reaction due to the rest-mass energy difference between both states. Finally, as the expansion rate of the universe became larger than the annihilation reaction rate, the global abundance of DM (its comoving number density) was frozen, and the probability of occurrence of an annihilation event became negligible.

In two well studied scenarios, the WIMP particles are such that they are their own anti-particle. For example in the case of a Kaluza-Klein excited state, arising naturally in compactified extra-dimensional models (like Universal Extra Dimensions or UED, in which all particles are able to propagate in the extra dimension provided they have enough energy), the WIMP is the lightest neutral gauge boson excited state, corresponding to the first excited state of the B boson or $B^{(1)}$, and as the standard model B boson, it is its own antiparticle (the B boson is a superposition of the photon of Quantum Electrodynamics and the Z boson of the weak interaction). Also, in the case of the MSSM (the minimal supersymetric standard model), considering conserved R-parity (and thus that superpartners are forbidden to decay to standard model particles), the WIMP is the lightest neutralino (i.e., the lightest mass eigenstate of the fermionic superpartners of the gauge bosons and the Higgses), and it is a Majorana fermion, and thus it is its own antiparticle.

In both these scenarios, the annihilation cross section is dependent only on the present day abundance of DM. This fact has been dubbed the WIMP miracle and is explained in what follows.

	\subsection{The WIMP miracle}

The WIMP miracle refers to the fact that, given that the WIMP is a thermal relic, its creation/annihilation cross section can be computed without needing extra model-dependent input (except for the fact that the cross section must not depend on the energy of the WIMP, i.e., the s-wave cross section must be dominant). The cross section then only depends on the present day DM density parameter $\Omega_{\chi,0}=(1-f_{bary})\Omega_{m,0}$, and the present day value of the Hubble parameter $h$ (that gives the current expansion rate). { As expected, $\Omega _{\chi,0}$ is inversely proportional to the cross section $<\sigma v>$.}

Obtaining this cross section is a standard derivation and it is done in, for example, Section 5.2 of \cite{EarlyUniverse}. For the value of the cross section we use

\begin{equation}
<\sigma v> = \frac{4 \times 10^{-27}}{\Omega_{\chi,0}h^{2}}[cm^{3}/s],
\end{equation}

\noindent which for the cosmology considered in this work corresponds to $<\sigma v> = 3.37 \times 10^{-26}[cm^{3}/s]$, and it is similar to the value used by \cite{HooperFinkbeinerDobler}.

Now we will motivate and explain two scenarios that give rise naturally to a thermal relic WIMP with the above-given cross section.

	\subsection{The KK DM particle}

In extra dimensional models (i.e., physical models that are built considering more than three spatial dimensions), the additional spatial degrees of freedom may be accessible to all or some of the particles. Also, these additional dimensions may be compactified differently and on different energy scales. If a standard model particle can propagate in this compactified (and thus finitely extended) dimensions, they will appear more massive in the four dimensional effective theory. In some cases, these excited states of standard model particles may have the desirable features of the WIMP. Here we will briefly discuss the lightest KK particle WIMP candidate and the extra-dimensional scenario in which it arises. We loosely follow the excellent review by \cite{HooperAndProfumo}.

The lightest KK particle { in the Universal Extra Dimensions (UED) framework \citep{Appelquist}} is the $B^{(1)}$, i.e., the first KK excited state of the B gauge boson (also called the KK-photon). This particle is charge neutral, and can annihilate with itself, and thus is a suitable WIMP candidate.

Some of the ways in which KK particles can interact with SM particles are through decays, annihilations and scatterings. For example, a heavier KK particle can decay into a lighter KK particle emitting standard model particles, until it has cascaded into the $B^{(1)}$ (also called $\gamma_{1}$, see Figure 4 of \cite{HooperAndProfumo}).

Also, more interestingly in view of the possible effects of WIMP annihilation on the IGM, two $B^{(1)}$s can annihilate to lepton pairs and photons (see Figure 16 of \cite{HooperAndProfumo}).

	\subsection{The Neutralino}

Supersymmetry is a particular symmetry property of the lagrangian of a physical theory, such that said lagrangian remains invariant if a supersymetric transformation is performed on the fields of the theory. The supersymmetric transformation is such that the change in a fermionic field is a bosonic field, and the change in a bosonic field is a fermionic filed. As in quantum field theory, the fields represent different particle contents (the particles are nothing more than the excited states of the fields), a supersymmetric theory naturally has one boson for each fermion present, and vice-versa.

The particle content of the MSSM (the Minimal Supersymmetric Standard Model) is the same as that of the SM, but for every SM particle, there is an associated supersymmetric partner (also, the Higgs field is different from the one in the SM, being two fields instead of just one). The naming scheme for the superpartners is to maintain the root of the name of the SM particle, but to change the end of the name with the suffix -ino in the case of fermionic superpartners of SM bosons and to add the prefix s- in the case of bosonic (scalar) superpartners of SM fermions. For example, the superpartners of the electron and the muon are the selectron and the smuon, and the superpartners of the photon and the Z boson are the photino and the Zino. The WIMP candidate of SUSY is the lightest neutralino, { as explained in the review by \cite{Jungman}}. 

The phenomenologies of KK particles (as they arise in UED) and of SUSY are very similar, and both of them could be tested, for example in particle colliders, by studying interactions above a certain energy scale. In the case of KK, this energy scale corresponds to the inverse of the radius of compactification of the extra dimmension, whereas in SUSY, it corresponds to the mass of the neutralinos and superpartners. Current collider lower limits on the mass scale of this new physics (for example the LHC limits) are on the order of the ~$100GeV$ and increasing. This mass limit will motivate the values that we adopt for the WIMP mass in the rest of this work. We use both a WIMP mass of $10GeV$ and of $1TeV$. 



\section{The DM Clustering}

As it was mentioned, the gravitational collapse process is hierarchical and self-similar, and so different levels of clustering are expected. In the following subsections we explain the prescriptions adopted to account for the DM clustering at all these levels, considering also the distribution of DM within a particular halo or subhalo. In {Section 3.1}, we follow the analytic approach, using the Press-Schechter formalism (\cite{PressSchechter}; \cite{MoWhite}; \cite{BondEtAl}; \cite{SMT}; \cite{LaceyCole} among others) to compute the number density of DM haloes per decade in mass. In {Section 3.2}, we account for the presence of substructure by considering the substructure mass function given by \cite{Giocoli}. In {Section 3.3}, we present the universal NFW density profile \citep{NFW}, that gives the DM distribution within each gravitationally bound halo, and explain how to calculate its parameters for different halo masses at different redshifts. Finally, in {Section 3.4}, we explain the effect that the gravitational collapse of the barionic matter (in particular, the formation of a supermassive black hole (SMBH)) can have on the DM density profile through adiabatic contraction.

	\subsection{The halo mass function}

We use this formalism, as presented in \cite{MoWhite}, but using the modified \cite{SMT} halo mass function and bias factor, calibrated with the GIF simulations \citep{Kauffmann}.

We consider the overdensity $\delta(x)$ defined as $\delta(x)=\frac{\rho(x)-\overline{\rho}}{\overline{\rho}}$, and the overdensity threshold for collapse $\delta_{c}(z)$, defined as the density contrast (in the linear approximation) required for an overdensity in a certain region to already have formed a collapsed object (\cite{NFW}, equation (A14)). We consider also the mass fluctuation $\sigma (R)$, defined as the standard deviation of the matter overdensity field when smoothed on a scale of size R. Finally, the dimensionless mass parameter is defined as $\nu = \frac{\delta_{c}(z)}{\sigma(R(M))}$, and it is a measurement of the mass of a collapsed region (DM halo) relative to the mass of the structures that have recently collapsed at redshift z.

We will use the \cite{SMT}  probability distribution $f(\nu)$. The mass fluctuation $\sigma(R)$ is computed following \cite{LaceyCole} and \cite{MoWhite}, starting from the power spectrum of matter overdensities, $P(k)$, at redshift z. { The latter is computed by applying the transfer function to the primordial power spectrum.} When calculating the mass function we adopt a primordial power spectrum of the form $P(k)=Ak^{n_{s}}$, where $A$ is a normalization and $n_{s}=0.9624$ \citep{Planck}.

The effects on the power spectrum due to the gravitational collapse of the DM on sub-horizon scales after the radiation-matter equality epoch and the Barion Acoustic Oscillations (BAOs) are encoded in the transfer function. We adopt the functional form given by Bardeen et al. (the BBKS transfer function), which ignores the BAOs.
				
	\subsection{The substructure mass function}

The second level of clustering of the DM corresponds to the self-bounded haloes that retain their identity within bigger haloes. The presence of these substructures is a natural consequence of the hierarchical structure formation picture, because haloes at all redshifts are formed by the aggregation of smaller parent haloes. As the DM is collisionless, DM haloes are significantly less disrupted than barionic matter during the merging processes, and thus subhaloes may survive up to the smallest scales (due to the negligible, but nonzero, free-streaming of the WIMPs). Many authors (for example \cite{KamionkowskyKoushiappas}; \cite{GiocoliPieri}; \cite{GiocoliTormen}; \cite{Taylor}; \cite{Pieri}) have investigated the clustering properties of the subhaloes by means of fitting an analytic form to the subhalo mass function obtained in high-resolution numerical N-body simulations and re-simulations. 

We use the substructure mass function proposed by \cite{Giocoli}, because it also considers the evolution of the substructure in time, due to the combined effects of gravitational heating and tidal stripping in the potential well of the main halo. These effects will tend to erode the subhaloes, which are the remnants of the haloes accreted by the host halo. As was found by these authors, the mass loss (to the smooth component of the main halo) of subhaloes can be approximated by an exponential decay of the subhalo mass on a characteristic timescale that is proportional to the dynamical time of the main halo. 

Explicitly, the unevolved subhalo mass function (that considers all subhaloes with the mass they had when they were accreted) is universal, and is given by:

 \begin{equation}
 \frac{dN}{dln(m_{v}/M_{0})} = N_{0}x^{-\alpha}e^{-6.283x^{3}} , x=\frac{m_{v}}{\alpha M_{0}},
 \end{equation}

\noindent
with $\alpha = 0.8$ and $N_{0}=0.21$. These values were calibrated using the GIF \citep{Kauffmann} and GIF2 \citep{Gao} simulations, as well as resimulations done by \cite{Dolag}. Here, $m_{v}$ represents the unevolved subhalo mass (the mass that the halo had when it was accreted) and $M_{0}$ represents the present-day host mass. { As a caveat of this formula we mention that the GIF simulation considered DM particles of $1.4\times 10^{10}M_{\odot}h^{-1}$, the GIF2 of $1.73\times 10^{9}M_{\odot}h^{-1}$ and the Dolag resimulations of $1.3\times 10^{9}M_{\odot}h^{-1}$, and therefore, for smaller masses (down to the free-streaming mass), the functional form of equation (2) may no longer hold.}

To account for time evolution, the authors find the following relation between the mass of a subhalo at time $t$ (given that it was accreted at time $t_{m}$), and its unevolved mass:

\begin{equation}
m_{sb}(t) = m_{v}exp\left[ -\frac{t-t_{m}}{\tau(z)}\right],
\end{equation}

\noindent
where $m_{sb}$ is the evolved mass and $\tau(z)$ is the characteristic mass loss time, given by

\begin{equation}
\tau (z) = \tau_{0} \left[\frac{\sigma(z,M_{0})}{\sigma(z=0,M_{0})} \right]^{-1/2}\left[\frac{H(z)}{H_{0}} \right]^{-1},
\end{equation}

\noindent
and $\tau_{0}=2.0Gyr$.

In this work, it was useful to approximate the accretion time of all subhaloes by the time at which the main halo collapsed, corresponding to a redshift $z_{coll}$, which we take as the redshift at which it acquired half of its mass (see Section 3.3). In terms of the look-back time to redshift z ($LBT(z)$), equation (3) can be approximated by:

\begin{equation}
m_{sb} = m_{v}exp\left[- \frac{LBT(z_{coll}) - LBT(z)}{\tau(z)} \right].
\end{equation}

Now, the required subhalo mass function is the one that includes the mass loss of subhaloes in time, so it should be $\frac{dN}{dm_{sb}}=\frac{dN}{dm_{v}} \frac{dm_{v}}{dm_{sb}}$. Expressing the subhalo mass function in terms of $m_{sb}$ instead of $m_{v}$, one obtains:

\begin{equation}
\frac{dN}{dm} = \frac{N_{0}}{m}x^{-\alpha}e^{-6.283x^{3}}, x = \frac{mK^{-1}(z,M_{h})}{\alpha M_{h}},
\end{equation}

\noindent
where $m$ is the present day (at redshift z), evolved mass of a subhalo, $M_{h}$ is the mass of the host halo and $K$ is defined by

\begin{equation}
K(z, M_{h}) = exp\left[- \frac{LBT(z_{coll}(M_{h})) - LBT(z)}{\tau(z)} \right].
\end{equation}

It can be seen that the behaviour of the subhalo mass function is that of a power law with an exponential cutoff, and that all mass scales are displaced as a function of time since halo formation. The displacement is such that a present day subhalo of mass $m$, corresponded to a subhalo of greater mass (by a factor $K$), and so its abundance is decreased because it is the abundance corresponding to the greater mass at the time of halo formation.

These last two equations give the number of subhaloes present in a halo of mass $M$ at redshift z, that have a mass between $m$ and $m+dm$, and so correspond to the subhalo mass function that we need. In the next section, we explore the smooth density profile of a particular halo, in order to account for the last level of the DM clustering hierarchy.

	\subsection{The NFW universal density profile}

The last hierarchy corresponds to how the DM is smoothly distributed within a particular halo or subhalo. This DM distribution appears to be universal, as found by \cite{NFW} using N-body simulations, and is given by the NFW density profile:

\begin{equation}
\rho(r) = \frac{\rho_{0}}{(r/r_{s})(1+r/r_{s})^{2}},
\end{equation}

\noindent
where $\rho_{0}$ is the characteristic density and $r_{s}$ is the scale radius. Thus, the NFW profile depends on only two parameters, { and for obtaining them, we first have to compute the redshift at which a halo with mass M (at current redshift z) was formed. The implicit definition of this redshift, called the redshift of collapse and denoted as $z_{coll}(M,z)$, is given by}

\begin{equation}
erfc\left(\frac{\delta_{c}(z_{coll}(M,z))-\delta_{c}(z)}{\sqrt{2\left( \sigma^{2}(0.01M,z) - \sigma^{2}(M,z)\right)}}\right) = 1/2.
\end{equation}

The choice of the mass of the small substructure (of $0.01M$) is motivated by comparisons to N-body simulations, { as mentioned in \cite{NFW}}.

{ Then}, we proceed to determine $\rho_{0}$ and $r_{s}$ using the following relations:

\begin{equation}
\rho_{0} = \rho_{crit}(z)\delta_{0}(M,z),
\end{equation}

\noindent
where $\rho_{crit}(z)$ is the critical density of the universe at redshift z, and $\delta_{0}(M,z)$ is a characteristic overdensity (or density contrast) that depends on the mass and redshift of the halo whose NFW profile we want to obtain.

Also,

\begin{equation}
r_{s} = \frac{r_{vir}(M,z)}{c(M,z)},
\end{equation}

\noindent
where $r_{vir}(M,z)$	is the virial radius of a halo of mass M at redshift z, corresponding the the radius at which the average density within is equal to the density required for the halo to be a self-gravitating virialized structure (in the case of the Einstein-De Sitter cosmology, this density is 200 times the mean density of the universe). $c(M,z)$ is the concentration parameter of the halo.

The virial radius is given by \cite{NFW} as:

\begin{eqnarray}
r_{vir}(M,z) = 1.63\times 10^{-2}\left(\frac{M}{h^{-1}M_{\odot}} \right)^{1/3}\times \nonumber\\ \times \left[\frac{\Omega_{m,0}}{\Omega_{m}(z)}\right]^{-1/3}(1+z)^{-1}h^{-1}kpc,
\end{eqnarray}

\noindent
where $\Omega_{m}(z)$ is the matter density	parameter at redshift z, and the virial radius is given in proper kpc.

The characteristic density is given by:

\begin{equation}
\delta_{0} = 3.41\times 10^{3}\Omega_{m}(z)\left(\frac{1+z_{coll}(M,z)}{1+z}\right)^{3}.
\end{equation}

Finally, the concentration parameter is implicitly given in terms of the characteristic density contrast by the relation:

\begin{equation}
\delta_{0}(M,z) = \frac{200}{3}\frac{c^{3}(M,z)}{\left[ln(1+c(M,z))-\frac{c(M,z)}{(1+c(M,z))} \right]}.
\end{equation}

Thus, the density profile of a particular halo or subhalo is, on average, completely specified by its mass and redshift. 

In the next section, we proceed to study the consequences of the collapse of the baryons on the density profile of the DM haloes, and obtain a modified density profile for the latter considering the growth of a central SMBH, a process that appears to be ubiquitous in galaxy formation.  

	\subsection{adiabatic contraction}

We consider the modification in the NFW DM density profile resulting from the gravitational collapse of baryons. Baryons behave dynamically differently from DM in their collapse process because they can heat and radiate away their energy, as they are not collisionless and have nonzero internal pressure. However, baryonic cooling will only take place in DM haloes with masses above a certain critical mass $M_{crit}$. A first estimate of this critical mass is the Jeans mass, such that the baryons in haloes with smaller masses are pressure-supported (see \cite{Barkana}). However, the Jeans criteria assumes that the DM perturbation from which the halo is formed is still in the linear regime, which is not the case for already virialized haloes, and therefore, \cite{Tegmark} consider a slightly different $M_{crit}$ criteria that accounts for the details of the collapse process. Finally, one could worry that the presence of DM itself may alter the $M_{crit}$, but as shown by \cite{RMF07b}, although the Jeans mass is altered, the $M_{crit}$ as considered by Tegmark changes by an $O(1)$ factor only (and this is considering LDM, because for the case of the massive WIMPs under study, it would be essentially unchanged). Therefore, in this work, we will consider that only those haloes (main and substructure) that attain a mass greater than $M_{crit}$, as given in the Figure 6 of \cite{Tegmark}, are able to undergo adiabatic contraction. For haloes of smaller mass, we will simply consider them to be of pure NFW form. (However, see \cite{Tanaka}, were the authors discuss a process that could allow for the formation of Intermediate Mass BHs at $z \sim 30$, which could generate adiabatic contraction, but only for very rare haloes).

{ In computing the adiabatically compressed density profile, different authors have considered different algorithms (e.g., \cite{Blumenthal}; \citep{Young}; \citep{SellwoodAndMcGaugh})}. { We consider the prescription given by \cite{NatarajanCrotonBertone}, that accounts} only for the adiabatic contraction due to the formation of a point mass (i.e., a SMBH) in the center of the potential well.

This algorithm assumes that the density profile is of the initial NFW form up to the radius of gravitational influence of the SMBH (defined as the radius up to which it is the main contributor to the contained mass), and that within this radius, the density profile is described by a power-law with a different (steeper) exponent. It also assumes (like the Blumenthal algorithm), that the original orbits of DM particles in the NFW profile were circular.

According to \cite{NatarajanCrotonBertone}, the value of the exponent of the density profile in the inner region is given by

\begin{equation}      
\gamma_{spike} = 2+\frac{1}{4-\gamma},
\end{equation}

\noindent
where $\gamma$ is the exponent in the inner region of the uncompressed DM density profile ($\gamma=1$ in the case of the NFW profile).

The derivation of the value of the exponent can be found in \cite{Quinlan}.

Finally, we need the maximum density and the radius of gravitational influence of the BH to have the complete density profile. For the maximum density, we use:

\begin{equation}
\rho_{DM}^{max} = \frac{m_{\chi}}{t_{spike}<\sigma v>},
\end{equation}

\noindent
where $m_{\chi}$ is the mass of the WIMP particle, and $<\sigma v>$ is the thermal relic annihilation cross-section (see {Section 2.1}). For $t_{spike}$, the time elapsed since the spike was formed, we { use $t_{spike} = LBT(z_{coll})-LBT(z)$}, i.e., the look-back time to the redshift at which the halo collapsed (defined in {Section 3.3}) as measured from the current redshift.

For the gravitational influence radius of the black hole, we use:

\begin{equation}
r_{BH} = 0.2\frac{GM_{BH}}{\sigma_{sph}^{2}},
\end{equation}

\noindent
where $M_{BH}$ is the black hole mass, and $\sigma_{sph}$ is the velocity dispersion of the spheroid component (the bulge in the case of haloes typical of late type galaxies).

For $\sigma_{sph}$, we use the observational relation between spheroid velocity dispersion and BH mass, as presented in \cite{Murray}:

\begin{equation}
M_{BH} = 1.5\times 10^{8}\left(\frac{\sigma_{sph}}{200[km/s]}\right)^{4}M_{\odot}.
\end{equation}

The mass of the black hole depends, on average, only on the mass of the host halo and on the fact that the host is a main halo or a subhalo. The relation between both masses is given by \citep{Lagos}:

\begin{equation}
log\left(\frac{M_{BH}}{M_{\odot}}\right) = 0.84 log\left(\frac{M_{halo}}{M_{\odot}}\right) - 2.1
\end{equation} 

\noindent
for the case of a main halo, and

\begin{equation}
log\left(\frac{M_{BH}}{M_{\odot}}\right) = 0.84 log\left(\frac{M_{halo}}{M_{\odot}}\right) - 2.9
\end{equation} 

\noindent
for the case of a subhalo.

Putting all the pieces together, the adiabatically compressed density profile of a halo has four different regimes, and is given by:

\begin{equation}
\rho_{DM}(r) = \rho_{NFW} (r) , r > r_{BH},
\end{equation}

\begin{equation}
\rho_{DM}(r) = \rho_{NFW}(r_{BH})\left(\frac{r}{r_{BH}}\right)^{-\gamma_{spike}} , r_{plateau} < r < r_{BH},
\end{equation}

\begin{equation}
\rho_{DM}(r) = \frac{m_{\chi}}{t_{spike}<\sigma v>} , 3r_{s} < r < r_{plateau},
\end{equation}

\begin{equation}
\rho_{DM}(r) = 0 , r < 3r_{s},
\end{equation}

\noindent
where $r_{plateau}$ is the radius at which the density of the spike equals the maximum attainable DM density, $r_{s}$ is the Schwarzschild radius of the BH { and the inner limit of $3r_{s}$ corresponds to the last stable orbit of a non-rotating BH}. In Figure 1, we show the adiabatically compressed density profile of a DM halo of mass $5\times10^{13} M_{\odot}$ at redshift $z=1$ (Typical of QSO systems). The four regimes are clearly distinguishable. The outer slope corresponds to the NFW $\gamma = 1$ and the inner slope corresponds to $\gamma_{spike} = \frac{7}{3}$.

\begin{figure}
\begin{center}
\includegraphics[width=0.5\textwidth]{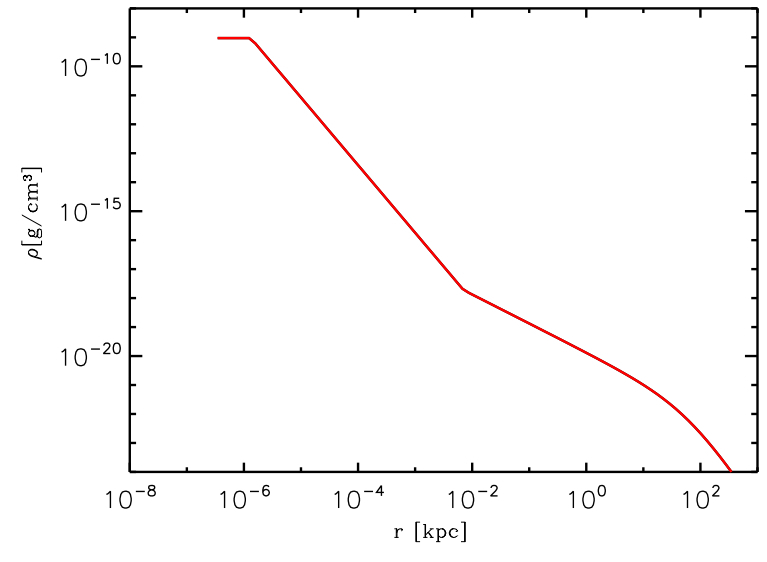}\\
\caption{{The adiabatically compressed density profile of a halo with mass $5\times10^{13}M_{\odot}$, at redshift $z=1$.}}
\label{nn}
\end{center}
\end{figure}	



\section{DM annihilation energy output}

Having considered the clustering properties of the DM at all scales and hierarchy levels, and having explained two particular DM WIMP candidates, motivating their expected masses and annihilation cross-sections, we can proceed to compute the expected energy output of annihilations in local structures (halos) and as an average for the whole IGM. In this section, we obtain the luminosity due to annihilations of individual DM haloes, considering the clustering analysed previously, and we also compute the annihilation energy rate injected to the IGM, per baryon, in the universe as a whole. We also mention the observational evidence of this annihilation energy injection process in satellite haloes and in the Milky Way (MW).  Finally, we { assess} the importance of the DM clumpiness in the energy output, comparing the case of a perfectly smooth universe with one with clustered DM.

First, let us consider the generic energy output. The reaction rate of the annihilation process is given by

\begin{equation}
\Gamma = n_{\chi}<\sigma v>,
\end{equation} 

\noindent
and thus, the variation of the number density of WIMPs can be calculated as:

\begin{equation}
\frac{dn}{dt} = \frac{1}{2}n_{\chi}^{2}<\sigma v>.
\end{equation}

As in the annihilation process the WIMPs are destroyed, we can consider that all their rest-mass is available as part of the energy output (actually, only a fraction of this energy will be available for affecting the environment, depending on the mechanism that couples the annihilation products and the IGM or halo medium). Then, the power density of the energy output can be written as:

\begin{equation}
\epsilon_{\chi} = \frac{1}{2}m_{\chi}c^{2}n_{\chi}^{2}<\sigma v>.
\end{equation}

This equation is completely general, and so it applies for haloes and also for the diffuse IGM. Let us now consider the power output in the different situations (we will call them local and global).

	\subsection{The smooth DM annihilation energy injection rate per baryon}

First we consider the global case, assuming a perfectly smooth DM density. This case is treated, for example, by \cite{MFP06}; \cite{RMF07}; and \cite{CLS}. As the universe may be infinite in spatial extent (as would be the case for flat, i.e. $\Lambda$CDM and Einstein-De Sitter, and Open cosmologies), the relevant quantity is the energy injected to the IGM per baryon. In terms of the present day WIMP number density, this is calculated as:

\begin{equation}
\dot{E}_{\chi}^{smooth} = \frac{1}{2}\frac{m_{\chi}}{n_{b,0}}c^{2}n_{\chi,0}^{2}(1+z)^{3}<\sigma v>f_{abs}(z),
\end{equation}   

\noindent
where $n_{\chi,0}$ is the present day DM number density, and is calculated as

\begin{equation}
n_{\chi,0} = \Omega_{\chi,0}\frac{\rho_{c,0}}{m_{\chi}},
\end{equation}

\noindent
where $\Omega_{\chi,0}=(1-f_{bary})\Omega_{m,0}$ is the present day DM density parameter, $\rho_{c,0}$ is the present day critical density of the universe and $m_{\chi}$ is the WIMP mass. Also, $n_{b,0}$ is the present day baryon density, and is calculated as

\begin{equation}
n_{b,0} = \Omega_{b,0}\frac{\rho_{c,0}}{\mu m_{H}},
\end{equation}

\noindent
where $\Omega_{b,0}$ is the present day baryonic density parameter (equal to $f_{bary}\Omega_{m,0}$), $m_{H}$ is the hydrogen mass and $\mu$ is the mean atomic weight of the baryon content. The mean atomic weight, considering a universe with only hydrogen and helium (a good approximation, particularly before star formation), can be computed as $\mu = f_{H} + 4f_{He}$, where $f_{H}$ is the fraction of hydrogen and $f_{He}$ is the fraction of helium by number. Assuming a value of $X = 0.74$ and $Y = 0.26$ (the hydrogen and helium fractions by mass, respectively), gives $f_{H} = 0.92$ and $f_{He}=0.08$. Thus, $\mu = 1.24$ will be used in this work.

Knowing the absorption fraction of the energy $f_{abs}(z)$, we can readily compute the energy injected to the IGM per baryon through DM annihilations. Different authors use various prescriptions for $f_{abs}(z)$. \cite{NatarajanAndSchwarz} consider that only the photons resulting as secondary annihilation products can inject energy to the IGM, and calculate the photon energy spectrum for different neutralino models explicitly. They consider that as the photons propagate, they lose energy, and compute the probability that they scatter off the IGM atoms. They find that $f_{abs}(z)$ is in the range of $0.1 - 0.2$ for $z<50$ (they do not consider higher redshifts). 

\cite{CLS}, compute $f_{abs}(z)$ for the neutralino (see also \citep{RMF07}), calculating first the number of photons and electrons produced per neutralino annihilation event, and they obtain $f_{abs}(z)$ in the range of $0.01-0.1$, being higher for higher z (and considering redshifts up to $z=1500$).

Other authors, like \cite{MFP06}, simply assume that $f_{abs}(z)=1$, and compute the maximal effects that their WIMP candidates can have on the IGM. We will consider that $f_{abs}$ is a constant that does not depend on redshift, and we will use the values $f_{abs}=$ 0.01, 0.1 and 1, in order to obtain results comparable to the different authors.

In {Figure 2}, we show the results obtained for the smooth energy injection per baryon, for two different WIMP masses ($1TeV$ and $10GeV$), as explained in the caption. We only consider the maximal absorption fraction ($f_{abs} = 1$), because for other values, the resulting curves should simply be rescaled by $f_{abs}$. It can be seen that the effect of a lower WIMP mass is, as expected, to boost the energy injection rate by a factor $\propto m_{\chi}^{-1}$.

\begin{figure}
\begin{center}
\includegraphics[width=0.5\textwidth]{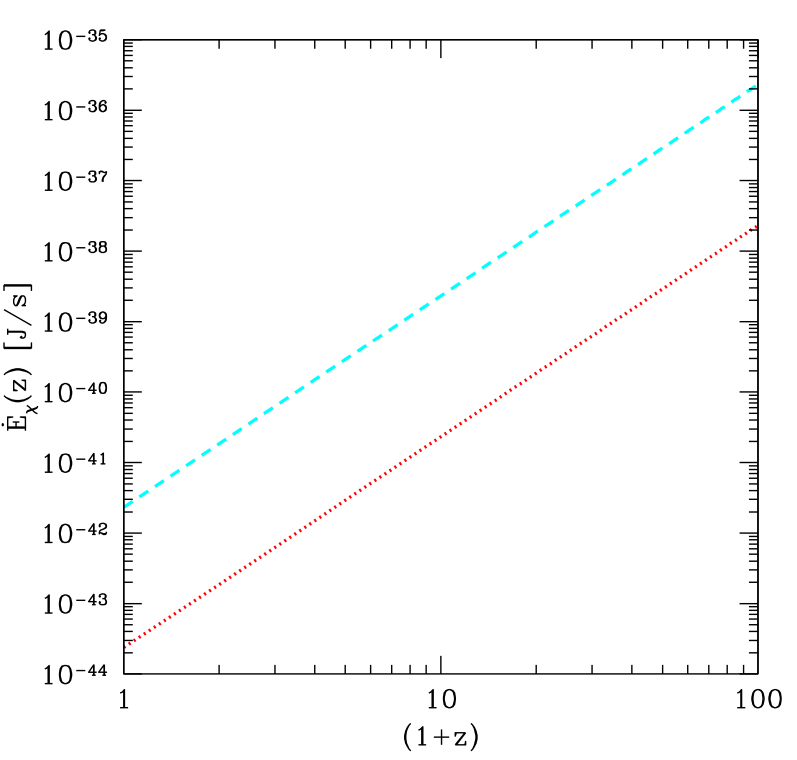}\\
\caption{{The smooth energy injection rate per baryon $\dot{E}_{\chi}(z)$, for different WIMP masses. The dotted curve corresponds to a WIMP mass of $1TeV$ and the dashed curve to a WIMP mass of $10GeV$.}}
\label{nn}
\end{center}
\end{figure}	

	\subsection{The DM annihilation luminosity of a halo}

Having considered the global case of the energy injection to the IGM, we now consider the energy output per unit time (or luminosity) of a particular halo of mass M at redshift z. Following \cite{NatarajanCrotonBertone}, from the general formula the energy output integrated on the volume of the halo is given by

\begin{equation}
L_{\chi} = \frac{<\sigma v>c^{2}}{2m_{\chi}}\int_{V}\rho_{DM}^{2}(x,t)d^{3}x,
\end{equation}

\noindent
where $\rho_{DM}$ is the complete DM density profile of the halo, accounting for the smooth main halo, the substructure and the adiabatic contraction in both (see Section 3.2 - Section 3.4). This luminosity can be decomposed into the main halo luminosity and the substructure luminosity such that:

\begin{equation}
L_{\chi} = L_{\chi , main} + L_{\chi , subs},
\end{equation}

\noindent
where 

\begin{eqnarray}
L_{\chi , main} = L_{\chi}(M_{main},z) = \nonumber\\ = \frac{<\sigma v>c^{2}}{2m_{\chi}}\int_{V}\rho_{DM}^{2}(r,M_{main},z)dV,
\end{eqnarray}

\noindent
and 

\begin{equation}
L_{\chi , subs} = L_{\chi , subs}(M_{main},z) = \int_{m_{free}}^{M_{main}}\frac{dN}{dm}L_{\chi}(m,z)dm.
\end{equation}

Note that in the above equations, $L_{\chi}(M,z)$ is the DM annihilation luminosity of a smooth, adiabatically compressed halo or subahlo, the only difference among them is that for the halo or the subhalo, the mass of the SMBH depends differently on the mass of the halo. It is important to mention that in order to account for the substructure luminosity, an integration of the substructure mass function multiplied by the luminosity of an individual subhalo in all the substructure mass range must be performed. This integration should be done between $m_{free}$, the free streaming mass of the WIMP (the minimum mass that a virialized DM halo can have), and $M_{main}$, the mass of the main halo (no substructure can be as massive as the main halo). The free streaming mass will depend naturally on the WIMP mass, as $m_{free} \propto m_{\chi}^{-3}$. Based on the analysis of the effects of the free streaming length on the matter power spectrum by \cite{Angulo}, we consider a free streaming mass of $m_{\chi} = 10^{-10}M_{\odot}$ for a WIMP mass of $m_{\chi} = 1TeV$, and a free streaming mass of $m_{\chi} = 10^{-4}M_{\odot}$ for a WIMP mass of $m_{\chi} = 10GeV$.
 
We explicitly compute the luminosity of DM haloes, in the $\Lambda$CDM cosmology, considering two different WIMP masses ($10GeV$ and $1TeV$) and the presence or absence of substructure and adiabatic contraction. { In {Figure 3}, we show the results in the form of the light to mass ratio ($L_{\chi}/M$) for the annihilation luminosity, as a function of halo mass.}

\begin{figure}
\begin{center}
\includegraphics[width=0.5\textwidth]{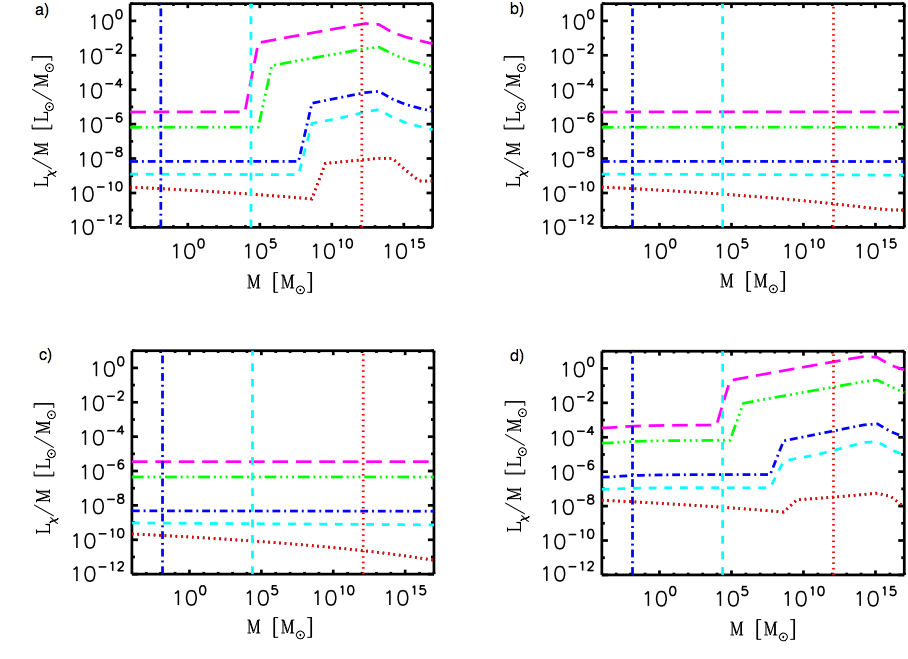}\\
\caption{{The { luminosity to mass ($L_{\chi}/M$) ratio} of the internal DM annihilation luminosity of individual haloes, as a function of halo mass (in the range from $10^{-4}$ to $10^{17}$ $M_{\odot}$). The different curves in the four plots correspond to different redshifts; in particular, the dotted curves correspond to $z = 0$, the short-dashed curves correspond to $z = 5$, the dot-dashed curves correspond to $z = 10$, the triple-dot-dashed curves to $z = 50$ and the long-dashed curves to $z = 100$. The different plots correspond to different clustering scenarios and WIMP masses; in particular, the top-left plot corresponds to the case with substructure and adiabatic contraction, with a WIMP mass of $1TeV$; the top-right plot corresponds to the case with substructure and $1TeV$ WIMP mass, but with no adiabatic contraction; the bottom-left plot corresponds to the case with a WIMP mass of $1TeV$ and { neither adiabatic contraction nor substructure}; and the bottom-right plot corresponds to the case with adiabatic contraction and substructure, but a WIMP mass of $10GeV$. In all the plots we also include the $\nu = 1$ mass values for the 3 lowest redshifts in the corresponding line-styles.}}
\label{nn}
\end{center}
\end{figure}	

It can be seen that, for the same halo mass, the luminosity of the halo increases with redshift. This is a result of the structure formation process, because a DM halo that formed earlier will be more concentrated and have a higher characteristic density. Thus, its annihilation rate, and in consequence its luminosity, will be increased. { Also, it can be seen for the cases that include adiabatic contraction, that for each redshift there is a discontinuity in luminosity at the mass corresponding to $M_{crit}$ (the critical mass for baryon cooling, as discussed in Section 3.4)}.
Also, it can be seen that the greatest effect on the luminosity comes from the WIMP mass. In accordance with what was found for the smooth case, $L_{\chi}/M$ is almost inversely proportional to $m_{\chi}$. There are variations to this proportionality in the cases with substructure and adiabatic contraction, for masses greater than $M_{crit}$, because the maximum attainable mass of the compressed spike is also dependent on $m_{\chi}$. It can be also seen that the second greatest effect on the luminosity comes from the presence of adiabatic contraction, { and that the presence of substructure contributes at most with an order $\sim$1 factor (at redshifts 5 and 10) to the luminosity when adiabatic contraction is absent, and thus it is the least significant contributor}. 

In order to assess the importance of DM annihilation as a possible feedback mechanism in haloes, we compare the obtained light to mass ratios for the annihilation luminosity with the typical mass to light ratios in haloes at different mass scales. For haloes with masses in the $10^{8}M_{\odot}-10^{15}M_{\odot}$ range, the logarithm of the mass to light ratio ($log_{10}(\Upsilon)$) at redshift zero (near universe) varies between 1.5 and 3, as shown by \cite{Eke} and also by \cite{Marinoni}. Also, $\Upsilon$ has a minimum for halo masses around $10^{12}M_{\odot}$, and it increases for lower masses due to SNe feedback and for higher masses due to AGN feedback (see \cite{Lagos} and also \cite{Mutch}). Thus, we note that the corresponding light to mass ratios for the haloes (${\Upsilon}^{-1}$) are much higher than the $L_{\chi}/M$ curves that we obtain for redshift zero. Furthermore, the minimum in $\Upsilon$ occurs for masses where the adiabatic contraction is already present, and therefore the $L_{\chi}/M$ is maximum, and thus both mass to light ratios correlate positively instead of negatively as it would be expected in the case of star-formation-quenching feedback. Therefore, we can conclude that the DM annihilation is irrelevant as a feedback mechanism. (However, see \cite{Ascasibar}.) It should be remembered that, although we plot halo luminosities up to haloes of mass of $10^{17}$, the most massive gravitationally bound haloes today (at $z = 0$), have masses of the order of $~10^{15}M_{\odot}$, so more massive haloes are virtually non-existent; and further more, the masses of the haloes that have a particular number density at a certain redshift (given by the characteristic dimensionless mass $\nu$), falls rapidly with redshift. This is the reason why even though the computed $L_{\chi}/M$ ratio of haloes with $M>M_{crit}$ at $z=50$ and $z=100$ is comparable to the near universe ${\Upsilon}^{-1}$, said haloes are, for practical purposes, negligible. For clarity, we also include in Figure 3, the lines corresponding to $\nu = 1$ at different redshifts.

	\subsection{The global, clumped, DM annihilation luminosity per baryon}

Having computed the DM annihilation luminosity of a particular halo of mass M at redshift z, we now return to the problem of computing the global energy injection rate to the IGM per baryon, but considering the clumpiness of the DM on all clustering scales. To do this, we simply consider that, knowing the luminosity of a particular halo (that already includes substructure and adiabatic contraction), and knowing the halo mass function (calculated in Section 3.1), the energy output rate density (or luminosity density) due to DM annihilations is given simply by:

\begin{equation}
\epsilon_{\chi}^{clumped}(z) = (1+z)^{3}\int_{m_{free}}^{\infty}L_{\chi}(M,z)\frac{dn}{dM}(M,z)dM,
\end{equation}

\noindent
where $m_{free}$ is the free-streaming mass, $L_{\chi}$ is the total DM annihilation luminosity (main halo plus substructure) of a DM halo of mass M at redshift z, $\frac{dn}{dM}$ is the previously calculated halo mass function (in comoving coordinates), and the factor $(1+z)^{3}$ is to convert the comoving density to a proper density. We consider $m_{max} = 10^{17}M_{\odot}$ since the number density of collapsed DM haloes with mass above this limiting mass is negligible.

Finally, the energy injection rate to the IGM per baryon can be directly computed as:

\begin{equation}
\dot{E}_{\chi}^{clumped}(z) = \frac{\epsilon_{\chi}^{clumped}(z)}{n_{b}(z)},
\end{equation}  	

\noindent
where $n_{b}(z)$ is the proper number density of baryons at redshift z (equal to $(1+z)^{3}n_{b,0}$ in terms of the present day baryon number density computed in Section 4.1).

It is important to mention that the energy injection rate to the IGM per baryon computed in this way, considers only the contribution of the DM haloes and not of the smooth component. In the case of a zero free-streaming mass, all the DM would be in haloes since there would be no minimum mass scale for the formation of collapsed structures. However, due to the non-zero free streaming mass, all haloes with masses less than that value, that should exist according to the halo mass function, are simply not allowed to form. As the clumped energy injection should be always greater than the smooth energy injection (computed in Section 4.1), and as the contribution due to DM haloes will be declining rapidly with increasing redshift after the truncation of low mass haloes due to the free-streaming limit becoming important, { we} use as the clumped DM injection rate the greater value among the one calculated in equation (46), and the one calculated in Section 4.1. 

The results obtained for the clumped energy injection rate per baryon, considering different clustering scenarios and WIMP masses, are shown in {Figure 4}. We only consider the maximal absorption fraction ($f_{abs} = 1$), because for other values, the resulting curves should simply be rescaled by $f_{abs}$.

\begin{figure}
\begin{center}
\includegraphics[width=0.5\textwidth]{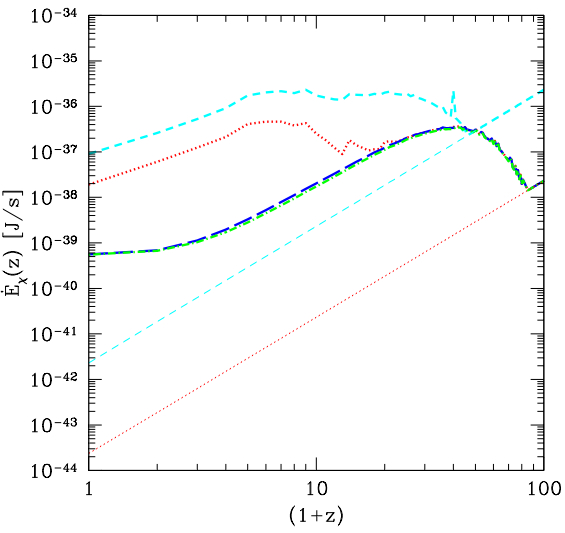}\\
\caption{{The clumped energy injection rate per baryon ($\dot{E}_{\chi}(z)$) as a function of redshift, for different scenarios, in thick lines. The dotted curve corresponds to the case with substructure and adiabatic contraction for a WIMP mass of $1TeV$, the long-dashed curve corresponds to the case with $1TeV$ WIMP mass and substructure but no adiabatic contraction, the dot-dashed curve corresponds to the case of $1TeV$ WIMP with { neither adiabatic contraction nor substructure}, and the short-dashed curve corresponds to a WIMP mass of $10GeV$ with substructure and adiabatic contraction. Also included as thin lines are the previous curves of Figure 2, corresponding to the smooth cases, for a $1TeV$ WIMP (dotted) and a $10GeV$ WIMP (short-dashed).}}
\label{nn}
\end{center}
\end{figure}		

It can be seen that the case considering a WIMP with mass of $10GeV$ is the one that gives the higher injection rate, in agreement with what was discussed above. { However, the curve for this case has a cutoff at $z \sim 50$ instead of at $z \sim 80$ like for the cases considering a $1TeV$ WIMP.} This can be understood as a consequence of the higher free-streaming mass (and thus, higher minimum mass that haloes can have), and the fact that the mass of the haloes that have a characteristic $\nu = 1$ (the haloes that have just recently collapsed) decreases with increasing redshift, such that with a higher free-streaming mass, less haloes are allowed to form at higher redshifts. { We can see that the presence of adiabatic contraction is also very important for boosting the annihilation energy injection. We also note that the curves considering adiabatic contraction have a secondary broad peak at lower redshift, besides the peak adjacent to the cutoff. This secondary peak is present due to the existence of $M_{crit}$, because for higher redshifts, the fraction of haloes that have $M>M_{crit}$ decreases due to the redshift dependence of the halo mass function, and so does the energy output in DM annihilations. { The curve for the case considering substructure but no adiabatic contraction is almost the same as that for the case considering only the NFW profile of the main halo, with neither adiabatic contraction nor substructure. The only slight difference occurs for intermediate redshifts and it amounts to an order $\sim$0.1 factor. The reason for this is that, as discussed in Section 4.2 (with respect to the interpretation of Figure 3), the contribution to the luminosity by the substructure, in the absence of adiabatic contraction, is of at most an order $\sim$1 factor for these redshifts, while at the same time, less massive haloes with less substructure are the most abundant.} 

In the next section, we obtain the clumpiness factor, C(z), implied by our energy injection calculations, and compare our results with those obtained by \cite{CLS}.	
	
	\subsection{The clumpiness factor}

The clumpiness factor is a useful tool when accounting for DM clustering in global DM annihilation energy injection computations, like the ones required for calculating the heating and ionization of the IGM due to DM (see Section 6.1 and Section 6.2). It simply relates the clumped energy injection to the smooth one by:

\begin{equation}
C(z) = \frac{\dot{E}_{\chi}^{clumped}(z)}{\dot{E}_{\chi}^{smooth}(z)}.
\end{equation}	

We show the values obtained for the clumpiness factor C(z) in {Figure 5}. The meaning of the curves are explained in the caption. 

\begin{figure}
\begin{center}
\includegraphics[width=0.5\textwidth]{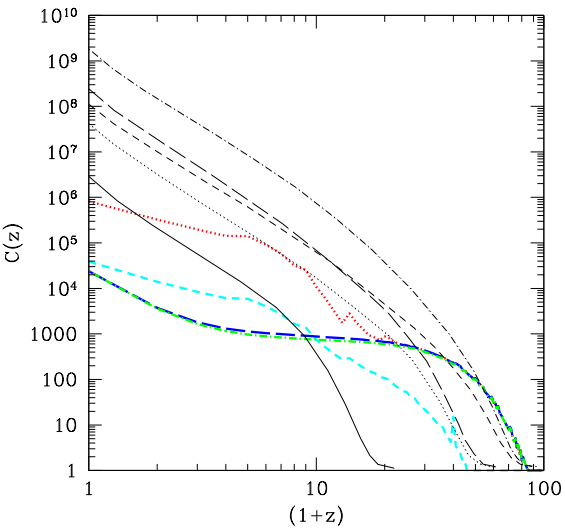}\\
\caption{{The clumpiness factor C(z) as a function of redshift. The thick curves correspond to our results. The line-styles of the thick curves are given as in {Figure 4}. For comparison, we also include the curves obtained by Cumberbatch, Lattanzi \& Silk (2010) for the case of an NFW profile, taken from the upper panel of their Figure 3, as thin curves. The line-styles of the thin curves correspond to different values of ($M_{min}/M_{\odot}$,$M_{cut}/M_{\odot}$). The solid line corresponds to ($10^{6}$,$10^{6}$), the dotted line corresponds to ($10^{-4}$,$10^{6}$), the long-dashed line corresponds to ($10^{-4}$,$10^{-4}$), the short-dashed line corresponds to ($10^{-12}$,$10^{6}$) and the dot-dashed line corresponds to ($10^{-12}$,$10^{-12}$). We refer the reader to the paper of the authors for further details.}}
\label{nn}
\end{center}
\end{figure}	

In general, it can be seen that for all the cases considered, the clumpiness is a decreasing function of redshift. This is reasonable, because as the redshift increases, the universe will tend to be more homogeneous and uniform, because the gravitational collapse and formation of structures would have had less and less time to occur. Also, for all the cases considered, the universe will be practically completely homogeneous by redshift ($z \sim 80$), except for the case with $10GeV$ DM WIMP, in which the universe becomes homogeneous at redshift ($z \sim 50$). Note that in this case, although $\dot{E}_{\chi}^{clumped}(z)$ is higher than for the other cases by almost an order of magnitude at redshift zero, $\dot{E}_{\chi}^{smooth}(z)$ is higher as well, and so the clumpiness C(z) is not. It can also be seen that in agreement with what was discussed before, the adiabatic contraction mechanism is important for low redshifts, and { the presence of substructure is irrelevant for increasing the clumpiness provided that the adiabatic contraction is present (for haloes and subhaloes with $M>M_{crit}$), as discussed in the previous sections.} 

We show also the clumpiness factors $C(z)$ obtained by \cite{CLS}, in Figure 5, using different line-styles. 
The two sets of curves can be readily compared as the definitions of the clumpiness factors are the same. There are, however, differences in the methods for computing them. In particular, \cite{CLS} compute the halo mass function using the P-S theory, starting from an analytic fit to $\sigma(M,z)$, and then consider that substructure and sub-substructure follow a pure power-law mass function with exponent $-2$. In general, it can be seen that we predict less clumpiness at low redshift, but comparable clumpiness at high redshift. The lower clumpiness predicted with respect to \cite{CLS} at low redshift can be understood because they also include sub-substructure (sub-subhaloes within subhaloes), which significantly enhances the annihilation output of the most massive haloes once they are already formed. Also, at intermediate redshift, our inclusion of adiabatic contraction increases our clumpiness signal, compensating for the lack of sub-substructure. At higher redshift, the contributions from both the substructure (and sub-substructure) and the adiabatic contraction are expected to be negligible, and therefore, any remaining difference would only be due to the use of slightly different prescriptions for computing the halo mass function.

We do not include Sommerfeld enhancement in our calculations, because it is highly dependent on unknown physics (the mass of the new force carrier and the resulting potential, as well as the physical scenarios in which this new force could arise), however, its effect is to directly modify $<\sigma v>$ by a constant factor. As we already mentioned in the Introduction, the presence of Sommerfeld enhancement is controversial and not supported by all observational evidence.

Having computed the DM annihilation luminosity generated within a DM halo for different scenarios, in the next section we study the luminosity received by a DM halo due to the rest of the universe.
	


\section{External DM annihilation energy injection on a halo}

On previous sections, we have considered particular WIMP models, calculated the DM clustering on all scales explicitly, and computed the DM annihilation luminosity generated within a particular DM halo, and the energy injection per baryon on the IGM. { In this section, we calculate the intensity ($Js^{-1}m^{-2}sr^{-1}$) of the average radiation field due to DM annihilations at different redshifts, considering our different clustering scenarios, and from this we obtain an upper bound on the power received by a halo per unit of halo mass ($L_{\odot}/M_{\odot}$), which can be compared with the luminosity to mass ratios due to annihilations inside the halo that were computed in the previous section. For this, we will use the annihilation luminosity of a halo that was obtained in Section 4.2, together with the cosmological radiative transfer formalism as presented in Section 2.1 of \cite{Haardt}. The power received from the radiation field by any halo will depend on its effective cross-section. For this cross section, we simply consider the number of electrons present in the halo times the Thompson cross-section. We later explain that this prescription, although crude, provides the desired upper bound on the received power per unit halo mass.
 
We now use the cosmological radiative transfer formalism in order to compute the intensity of the radiation field and the corresponding bound on the received power per unit halo mass.}

\subsection{The radiative transfer of the DM annihilation output.}

In section 2.1 of \cite{Haardt}, the solution to the problem of radiative transfer, in terms of the isotropic specific intensity, in the context of a dynamic cosmological background was presented. The authors considered the radiative transfer equation,

\begin{equation}
\left(\frac{\partial}{\partial t}-\nu \frac{\dot{a}}{a}\frac{\partial}{\partial \nu}\right) J =-3\frac{\dot{a}}{a}J-c\kappa J + \frac{c}{4\pi}\epsilon,
\end{equation}

\noindent with formal solution

\begin{equation}
J(\nu_{0},z_{0})=\frac{1}{4\pi}\int_{z_{0}}^{\infty}dz \frac{dl}{dz}\frac{(1+z_{0})^{3}}{(1+z)^{3}}\epsilon (\nu,z)e^{-\tau_{eff}(\nu_{0},z_{0},z)},
\end{equation}

\noindent where J is the specific intensity (in units of $Js^{-1}m^{-2}Hz^{-1}sr^{-1}$), $\epsilon$ is the specific emissivity, $\nu$ is the redshifted frequency of the radiation, $\frac{dl}{dz}$ is the distance travelled by a single photon of radiation per redshift interval and $\tau_{eff}$ is the effective optical depth for the radiation travel.

In our case, because we are considering a model-independent WIMP scenario, we do not have a specific spectral energy distribution (SED) for the radiation resulting from the annihilation events. Therefore, instead of using a frequency dependent specific intensity, we modify this formalism to consider an intensity integrated in frequency. In this case, the emissivity will be the usual power density, and we have to include an extra $(1+z)$ factor to account for the redshifting of the energy of all photons (independently of the details of the SED in frequency). Thus, we write

\begin{equation}
J(z_{0})=\frac{1}{4\pi}\int_{z_{0}}^{\infty}dz \frac{dl}{dz}\frac{(1+z_{0})^{4}}{(1+z)^{4}}\epsilon _{\chi}(z)e^{-\tau_{eff}(z_{0},z)}.
\end{equation}

Now, we give the prescriptions used for the terms included in this equation.

For the line element, we use 
\begin{equation}
\frac{dl}{dz}=c\frac{dLBT(z)}{dz},
\end{equation}

\noindent where $LBT(z)$ is the look-back time to redshift z.

{ We also compute the emissivity of the annihilation luminosity as

\begin{equation}
\epsilon_{\chi}(z)=\frac{1}{2}m_{\chi}c^{2}n_{\chi,0}^{2}(1+z)^{6}<\sigma v>C(z),
\end{equation}

\noindent which is the same emissivity as the one given in equation (35), when the definition of the clumpiness (given in Section 4.4) is considered.}

For the effective optical depth, we use

\begin{eqnarray}
\tau_{eff}\left(z_{0},z\right)=\int_{m_{free}}^{\infty}\int_{z_{0}}^{z}dMdz'\left(\frac{dl}{dz'}\right)\frac{dn}{dM}\left(M,z'\right)\times\\
\nonumber \times (1+z')^{3}\sigma_{bary}(M),
\end{eqnarray}

\noindent where we specify $\sigma_{bary}$ as follows.

The baryonic cross-section of a halo to the DM annihilation radiation was computed as

\begin{equation}
\sigma_{bary}(M)=\sigma_{T}f_{bary}\left(\frac{M}{m_{p}}\right)\left(1-\frac{Y_{p}}{2}\right),
\end{equation}

\noindent where $f_{bary}=\frac{\Omega_b}{\Omega_m}=0.17$ is the fraction of baryons by mass, $Y_{p}=0.26$ is the primordial abundance of He by mass and $\sigma_{T}$ is the Thompson cross-section. This corresponds to considering the total cross section of a DM halo as the number of electrons present in the halo times the Thompson cross-section. Although this prescription is crude, it is also completely model-independent. { The implicit assumption here is that photons of the radiation are energetic enough that bound electrons can be considered as free (i.e. that their energies are $\gtrsim$ 0.1-1$keV$ for metal-free gas, and $\gtrsim$ 1-10$keV$ for metal-enriched gas)}; and also that they are soft enough so that their energies are small with respect to the rest-mass energy of the electrons (to avoid pair creation). The first assumption is certainly true, whereas the second assumption seems dubious because the mass of an electron is on the $MeV$ scale whereas the mass of a WIMP is on the $GeV$-$TeV$ scale. However, one should keep in mind that the radiation background received by the halo is not necessarily the same radiation that was obtained as a secondary product of the annihilations, because the photons have been redshifted as they travelled through the universe, and also, they could have been scattered off other haloes before. { Also, this prescription provides an (approximate) upper bound on the power absorbed by a halo from the radiation intensity field. To see this, first we mention that for all considered redshifts, we are always in the optically thin regime ($\tau_{eff} \ll 1$), and thus, we can ignore the exponential suppression term in the computation of $J$. This is in agreement with the results obtained by \cite{Zdziarski} for the optical depth for cosmological propagation of radiation in the IGM, computed performing a detailed analysis depending on the energy of the radiation photons, which we do not consider here because we do not have a particular SED for the radiation as we intend to keep the discussion model independent. In any case, ignoring the exponential suppression, the absorbed power will be simply proportional to the cross-section of the halo, and therefore, the use of the Thompson cross-section as the cross-section of each individual electron amounts to obtaining an upper bound on the absorbed power because for higher energy photons, the correct Klein-Nishina correction factor only decreases the cross-section (when neglecting pair production).}

{ Thus, considering the prescriptions given above, the terms that enter in equation (40) are now fully specified.

We note that our expression for the intensity considers the average clumpiness, and therefore it neglects the existence of excess clustering about haloes of a given mass. In principle, one may expect this clustering to play a role in the computation of the specific intensity. However, we find that if one accounts for this excess clustering (that depends on the correlation function and the bias factor) by computing the intensity as centred about a halo of mass M, one obtains almost the same result than for the case considering the average clumpiness up to a small difference of $\sim$10\%. }

Now, given the specific intensity of the radiation field, we compute the flux that is incident on the surface of the halo, and the energy that is absorbed by it. By surface, we mean that we assign a radius to the halo that corresponds to $R=\sqrt{\frac{\sigma_{bary}(M)}{\pi}}$.

{ Considering that J is isotropic, the} flux incident on any surface element of the halo can be written as

\begin{equation}
F=\pi J.
\end{equation}

Finally, the power received by a halo will be given by

\begin{equation}
L^{ext}_{\chi}=4\pi R^{2} F,
\end{equation}

\noindent where $R$ is the radius of the halo. This can also be written as
\begin{equation}
L^{ext}_{\chi}=4\pi \sigma_{bary}(M) J.
\end{equation} 

We see that the final result is proportional to $\sigma_{bary}(M)$, and therefore, for the reasons already discussed, it should only be considered as an upper bound on the { received power. Also, we are implicitly assuming that the entirety of the annihilation energy that is produced in DM haloes is available to propagate through the universe and thus contributes to the intensity of the radiation field. This need not be the case, as some fraction of this radiation may be absorbed in the same haloes where it is produced. In fact, in the computations of Section 6 it is assumed that a fraction $f_{abs}$ of the annihilation luminosity is instantaneously absorbed by the IGM. This assumption, however, is consistent with the computation of an upper bound to the received power.}

\begin{figure}
\begin{center}
\includegraphics[width=0.5\textwidth]{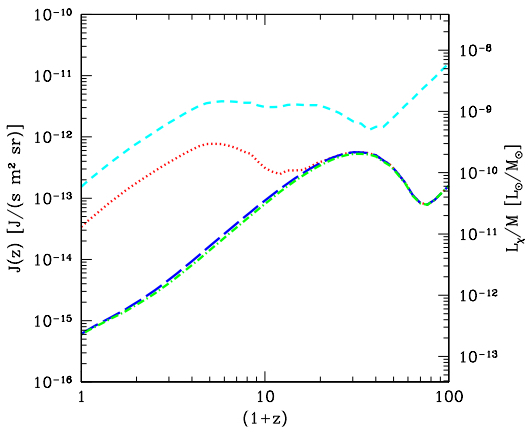}\\
\caption{The intensity of the radiation field due to DM annihilations ($J(z)$) and its corresponding luminosity to mass ($L^{ext}_{\chi}/M$) ratio, as a function of redshift, for different DM clumpiness scenarios. The line-styles of the curves are given as in Figure 4.}
\label{nn}
\end{center}
\end{figure}

{
In Figure 6, we show our results for the intensity of the radiation field as a function of redshift, and the corresponding light-to-mass ratio (defined as $L^{ext}_{\chi}/M$), for the different clustering scenarios and WIMP masses under consideration (as discussed in Section 4). It can be seen that the WIMP mass has the greatest effect on the intensity, being higher for the $10GeV$ WIMP by about an order of magnitude compared to the $1TeV$ case. The presence of adiabatic contraction also has a significant impact at lower redshifts, resulting in a $\sim$2 orders of magnitude increase with respect to the cases where it is absent, while the presence of substructure has an almost negligible effect. The intensity curves have an overall increasing behaviour with redshift that is due to the cosmic expansion and to the redshifting of light, although it is not a monotonous trend and shows two local maxima or humps. As the redshift increases, for the cases that consider adiabatic contraction, there is a hump followed by a decrease due to the fact that the population of haloes with masses above $M_{crit}$ (the mass threshold for baryonic cooling) is significant only at later cosmic times. There is another hump at higher redshifts after which the curves decrease their intensity due to the scarcity of collapsed haloes with masses bigger than the free streaming mass. These features are similar to those of the energy injection rate per baryon curves shown in Figure 4. 

Because the power that is received by a halo from the radiation intensity field is proportional to the halo cross-section, and because this is in turn proportional to the number of baryons in the halo and therefore to its mass, the light-to-mass ratio ($L_{\chi}/M$) here obtained is independent of halo mass, and it is only a function of redshift. This dependance is also shown in Figure 6, and it is seen that the ratio calculated here is lower than the corresponding ratio obtained for the annihilation luminosity produced within DM haloes (shown in Figure 3) by at least an order of magnitude, for all masses at the same redshifts, the smaller difference occurring for the case of haloes with mass of $\sim 10^{8}M_{\odot}$ at $z=0$ and with a $1TeV$ WIMP and adiabatic contraction. Therefore, the DM annihilation power received from the radiation intensity field is not relevant as a feedback mechanism in haloes.
}



\section{Global IGM heating, ionization and 21 cm signal due to DM}

Different authors have studied the effects of DM annihilations (or decays) on the global properties of the IGM, due to their extra energy injection (for example, see \cite{Pierpaoli}; \cite{MFP06}; \cite{RMF07}; \cite{VFMR07}; \cite{NatarajanAndSchwarz}; \cite{CLS}; {  \cite{Furlanetto}; \cite{Chuzhoy}} among others). In this section, we study the effects on the IGM of the energy injection found on Section 4.3. We divide the section into different subsections, each corresponding to one of the different physical quantities of interest that are affected by this extra energy injection. We explain the formalism and the definition of the quantities, as well as how to compute them, in the beginning of the corresponding subsections, and then we show the results we obtained and compare them with different authors. The first two sections (the ionization fraction and the IGM temperature), refer to general properties of the IGM, whereas the next sections have to do with the signal of this extra-heating on the 21 cm line of the hyperfine structure of neutral hydrogen (HI). 

We start, in the next subsection, with the computation of the IGM ionization fraction.

	\subsection{The IGM ionization fraction}

The ionization fraction ($x_{e}$) is defined as the ratio of free electrons to hydrogen atoms $x_{e} = \frac{n_{e}}{n_{H0}+n_{H+}}$. Following \cite{VFMR07}, the variation of the ionization fraction is given by:

\begin{equation}
\frac{dx_{e}}{dz} = \frac{1}{H(z)(1+z)}\left[R_{norad}(z) - I_{norad}(z) - I_{\chi}(z)\right],
\end{equation}

\noindent
where $I_{norad}$ and $R_{norad}$ are the ionization and recombination rates per baryon { in the absence of annihilating DM and ionizing radiation sources}, and $I_{\chi} = \chi_{i}(z)\frac{\dot{E}_{\chi}}{E_{0}}$ is the contribution to the ionization rate due to the energy injection of DM annihilation. $\dot{E}_{\chi}$ is the energy injection rate per baryon computed in Section 4.3, $E_{0}$ is the ionization energy of Hydrogen (13.6 eV), and $\chi_{i}(z)$ is the ionization efficiency and is given by $\chi_{i}(z) = \frac{1-x_{e}}{3}$ { \citep{Shull}}. The factor $\frac{1}{H(z)(1+z)}$, is simply a change of variables, because the differential of cosmic time $dt$ and the redshift differential $dz$ are related by $dt = \frac{-dz}{H(z)(1+z)}$. 

We define the DM annihilation contribution to the ionization fraction as $\delta x_{e}(z) = x_{e}(z) - x_{e}^{norad}(z)$, where $x_{e}^{norad}$ is the ionization fraction in a scenario with no ionization sources. It is important to mention that this ionization fraction is a consequence of the physics of recombination from the epoch of the last scattering surface, and is simply the relic free-electron abundance from the freeze-out of the recombination reaction. Thus, the $x_{e}^{norad}$ does not account for the reionization of the universe by Poulation III stars, that according to Planck 2013 \citep{Planck}, occurred { around} redshift $\sim$11. With this definition of $\delta x_{e}$, we have:

\begin{equation}
\frac{d\delta x_{e}}{dz} = \frac{-I_{\chi}(z)}{H(z)(1+z)}.
\end{equation}

We solve this differential equation, using the $x_{e}^{norad}$ given in Figure 3 of \cite{MFP06}, by the Euler method, using as boundary condition that $\delta x_{e}(z=1000) = 0$ (at redshift $z>1000$, as we are near the recombination epoch, the ionization fraction is near 1 and is almost entirely due to the standard cooling scenario). 

We show the obtained ionization fraction ($x_{e}(z)$) curves in Figure 7 and in Figure 8. The first figure shows the $x_{e}(z)$ curves obtained for the different DM clustering scenarios and WIMP masses, but considering the clumped DM case, and with an absorption fraction of $f_{abs} = 1$ (more details on the caption of the figure). The second figure shows the ionization fractions, assuming a WIMP mass of $1TeV$ and substructure and adiabatic contraction, but for different values of the absorption fraction ($f_{abs}=0.1$ or $0.01$), or simply not considering DM clustering at all, and using the homogeneous DM energy injection computed previously.

\begin{figure}
\begin{center}
\includegraphics[width=0.5\textwidth]{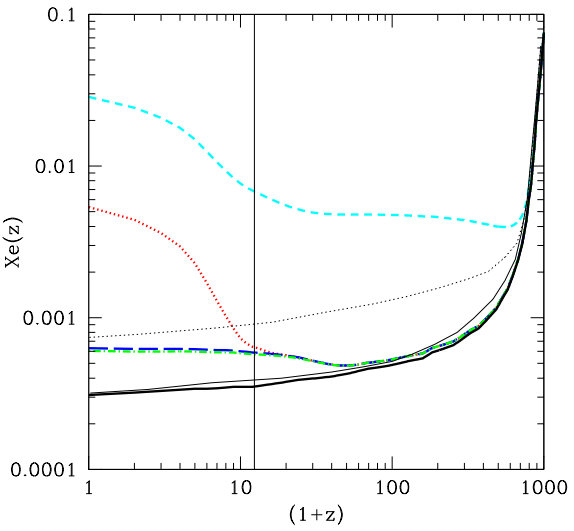}\\
\caption{{The ionization fraction of the IGM for different cases, assuming a maximal absorption fraction and the clumpiness of the DM. Our results are shown as thick lines. The different line-styles of the thick lines correspond to: (short-dashed) a WIMP mass of $10GeV$ with adiabatic contraction and substructure, (dotted) a WIMP mass of $1TeV$ with adiabatic contraction and substructure, (long-dashed) a WIMP mass of $1TeV$ with substructure but no adiabatic contraction, (dot-dashed) a WIMP mass of $1TeV$ with { neither adiabatic contraction nor substructure}, and (solid) the standard cooling scenario, with no DM energy injection. For comparison, we also include the results obtained by Mapelli, Ferrara \& Pierpaoli (2006) for the case of the $100GeV$ SUSY neutralino, taken from the bottom panel of Figure 3 of the authors and we show them as thin lines. The thin solid line corresponds to $<\sigma v> = 2 \times 10^{-26}cm^{3}s^{-1}$ and the thin dotted line corresponds to $<\sigma v> = 10^{-24}cm^{3}s^{-1}$. Also included is the line corresponding to $z = 11.35$, that indicates when the effects of PopIII stars { likely} become important.}}
\label{nn}
\end{center}
\end{figure}	

\begin{figure}
\begin{center}
\includegraphics[width=0.5\textwidth]{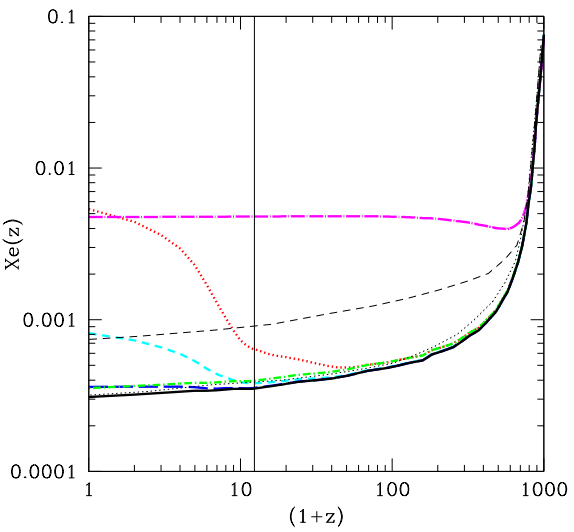}\\
\caption{{The ionization fraction of the IGM for different cases, { considering substructure and adiabatic contraction but with different absorption fractions,} or no DM clumpiness at all. Our results are shown as thick lines. The different line-styles of the thick lines correspond to: (dotted) the standard clustering scenario considering both substructure and adiabatic contraction with a $1TeV$ WIMP and $f_{abs}=1$, (short-dashed) the same clustering scenario but considering $f_{abs}=0.1$, (long-dashed) the same clustering scenario but with $f_{abs} = 0.01$, (short-dot-dashed) the same WIMP mass but considering no DM clumpiness, (long-dot-dashed) a $10GeV$ WIMP with no DM clumpiness, and (solid) the standard cooling scenario, with no DM energy injection. For comparison, we also include the results obtained by Mapelli, Ferrara \& Pierpaoli (2006) for the case of the $100GeV$ SUSY neutralino, taken from the bottom panel of Figure 3 of the authors and we show them as thin lines. The thin solid line corresponds to $<\sigma v> = 2 \times 10^{-26}cm^{3}s^{-1}$ and the thin dotted line corresponds to $<\sigma v> = 10^{-24}cm^{3}s^{-1}$. Also included is the line corresponding to $z = 11.35$, that indicates when the effects of PopIII stars { likely} become important.}}
\label{nn}
\end{center}
\end{figure}	

It can be seen from Figure 7 that the ionization fraction evolves very differently from the case with no DM energy injection, as long as the DM clumpiness is considered. Also, for the case of a WIMP with a mass of $10GeV$, the ionization fraction can be increased up to $x_{e} \sim 0.03$ at redshift zero, whereas for the other cases that consider a $1TeV$ WIMP, this partial re-ionization can only bring $x_{e}$ to $\sim 0.005$ at redshift zero, hence leading to a difference of one order of magnitude of the final ionization fraction. Note, however, that the computation of the ionization fraction does not account for the PopIII stars (for a review, see Section 4 of \cite{Barkana}), that are thought to have re-ionized the universe around redshift $z\sim 11$ according to the results by Planck \citep{Planck}; thus, { these} results are only reliable down to { approximately that redshift, and consequently we consider such limit in the plots by adding it as a vertical line}. Note, nonetheless, that we should keep in mind that the ionization process is by no means an instantaneous process, and that this redshift corresponds only to the best fit for the six-parameter base $\Lambda CDM$ model under the assumption of instantaneous reionization (as explained in \citep{Planck}), { which approximately corresponds to the redshift at which the neutral hydrogen regions had a volume filling factor of $0.5$}, but that large regions of neutral Hydrogen may have still been present up to { $z\sim 7$}, with a volume filling factor of $\gtrsim 0.1$ (see section 5.1 of \cite{RMZ}). Also, we note that as discussed by \cite{RMZ} and by \cite{Ricotti}, the reionization may have been caused not only by the first stars, but also by the X-ray emission from accretion onto early Black Holes.

{ At $z \sim 10$, where the effects of DM annihilation could in principle be observed, the ionization fraction could be raised to $6 \times 10^{-4}$, corresponding to a factor of 2 increase with respect to the adiabatic cooling case. In the case of a $10GeV$ WIMP, the ionization fraction at $z\sim 10$ could be increased to $8 \times 10^{-3} $. }

It can also be seen in { Figure 7} that adiabatic contraction is considerably less important than WIMP mass in determining the evolution of the ionization fraction, and that it becomes relevant only for $z \sim 10$. { Also, for a $1TeV$ WIMP, the case with neither substructure nor adiabatic contraction results in $x_{e} \sim 0.0006$ at $z = 0$, being almost equivalent (up to an order $\sim0.1$ factor) to the case with substructure and no adiabatic contraction. Therefore, in agreement with what was mentioned in previous sections, the presence of substructure has little to no effect on the evolution of the IGM kinetic temperature.}

From Figure 8, it can be seen that even without considering DM clumpiness, the case with a $10GeV$ WIMP considerably modifies the ionization fraction of the universe. It has the effect of raising the relic ionization at $z \sim 10$ by a factor of $\sim 20$. Also, the evolution of the ionization fraction is particularly sensitive to the absorption fraction $f_{abs}$, such that { for the $1TeV$ WIMP case}, reducing $f_{abs}$ from $1$ to $0.1$ results in a completely different evolution for $z \lesssim 10$ (and in a ionization fraction at $z=0$ an order of magnitude lower). The cases with $f_{abs} = 0.1$ and no clumpiness give evolutions of the ionization fraction that are very similar to the case were no DM energy injection is present.

	\subsection{The IGM kinetic temperature}

We now consider the kinetic temperature of the IGM. Following \cite{VFMR07}, the change in this temperature is given by:

\begin{eqnarray}
(1+z)\frac{dT_{k}}{dz} = 2T_{k} + \frac{l_{\gamma}x_{e}}{H(z)(1+Y+x_{e})}(T_{k}-T_{CMB}) - \nonumber
\end{eqnarray}
\begin{equation}
- \frac{2\chi_{h}\dot{E}_{\chi}}{3k_{b}H(z)(1+Y+x_{e})},
\end{equation}

\noindent
where $l_{\gamma} = \frac{(8\sigma_{T}a_{R}T^{4}_{CMB})}{(3m_{e}c)}$ is the analogue to a thermal conductivity between the CMB and the IGM, $\chi_{h}$ is the heating efficiency and is given by $\chi_{h} = \frac{1+2x_{e}}{3}$ and $Y$ is the helium fraction by mass (taken as $Y = 0.26$).

It is expected (see for example \cite{CLS}) that the IGM kinetic temperature tracks the CMB temperature at high redshifts, so we used the CMB temperature at redshift $z=1000$ as the boundary condition for the IGM kinetic temperature, and solved the differential equation by the Euler method. The interpretation of this cooling equation is straightforward, as the first term on the right hand side corresponds to the adiabatic cooling of the IGM in absence of heat exchange or energy injection, and proceeds as $T \propto (1+z)^{2}$. The second term corresponds to the heat exchange between the CMB and the IGM, and the third term allows for the extra heating due to the DM annihilation energy injection. 

We show the obtained IGM kinetic temperature curves in Figure 9 and in Figure 10. The first figure shows the temperatures obtained for the different DM clustering scenarios and WIMP masses, but considering the clumped DM case, and with an absorption fraction of $f_{abs} = 1$ (more details on the caption of the figure). The second figure shows the temperatures, assuming a WIMP mass of $1TeV$ and substructure and adiabatic contraction, but for different values of the absorption fraction ($f_{abs}=0.1$ or $0.01$), or simply not considering DM clustering at all, and using the homogeneous DM energy injection computed previously.

\begin{figure}
\begin{center}
\includegraphics[width=0.5\textwidth]{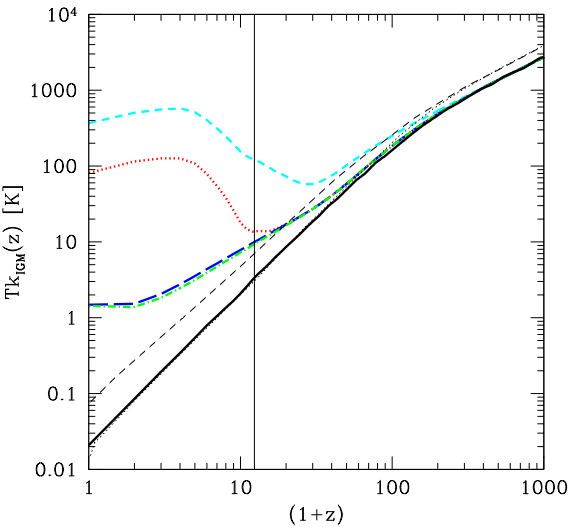}\\
\caption{{The kinetic temperature of the IGM for different cases, assuming a maximal absorption fraction and the clumpiness of the DM. For comparison, we also include the results of Mapelli, Ferrara \& Pierpaoli (2006), taken from the top panel of Figure 3 of the authors. The different line-styles are assigned in the same way as in Figure 7. The $z=11.35$ line is also included.}}
\label{nn}
\end{center}
\end{figure}	

\begin{figure}
\begin{center}
\includegraphics[width=0.5\textwidth]{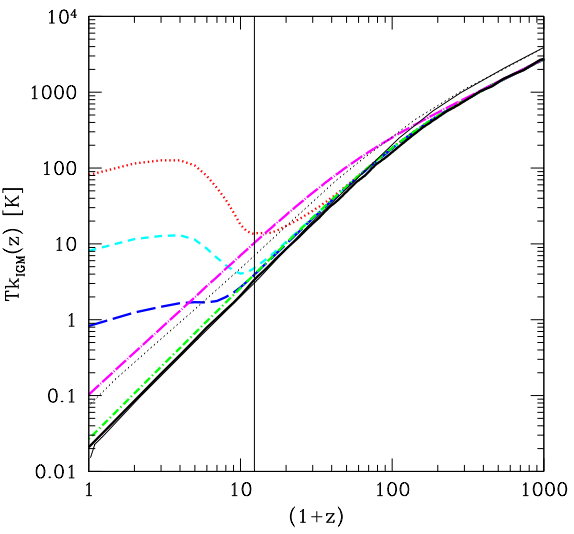}\\
\caption{{The kinetic temperature of the IGM for different cases, considering substructure and adiabatic contraction but with different absorption fractions, or no DM clumpiness at all. For comparison, we also include the results of Mapelli, Ferrara \& Pierpaoli (2006), taken from the top panel of Figure 3 of the authors. The different line-styles are assigned in the same way as in {Figure 8}. The $z=11.35$ line is also included.}}
\label{nn}
\end{center}
\end{figure}	

In can be seen on Figure 9 that for all the cases that consider the clumpiness of the DM (assuming $f_{abs}=1$), the cooling is very different from the standard adiabatic scenario. Significant re-heating occurs in the case of a $10GeV$ WIMP, where the IGM may be heated even to temperatures of $ \lesssim 1000 K$ by redshift zero (although, remember that this cooling scenario does not account for energy injection due to stellar components and early black holes, and so it is reliable only down to redshift $z \sim 10$). Also, adiabatic contraction is more important for the cooling history than substructure, and the extra heating due to this compression, { for the case of a $1TeV$ WIMP,} can amount to almost two { orders} of magnitude of difference by redshift zero, but the difference is relevant only for $z \lesssim 10$. { At $z\sim 10$, the IGM may be heated up to temperatures of $10K$ for the case of a $1TeV$ WIMP, corresponding to an increase by a factor of 10 with respect to the adiabatic cooling scenario. For a $10GeV$ WIMP, the temperature at $z\sim 10$ can be raised to $200K$.} 

Also, it can be seen on Figure 10 that the absorption fraction is critical for determining the particular cooling history, and that the low-redshift heating due to DM annihilations can vary enormously with different choices for $f_{abs}$. It can be seen that, in general, in the cases considering adiabatic contraction, the IGM kinetic temperature reaches a quasi-stationary state, and the final temperature attained is almost proportional to the absorption fraction.

We also include the results for the ionization fraction and IGM kinetic temperature obtained by \cite{MFP06}, in order to compare them with our results. It can be seen that for the cases	with no DM clumpiness, we reproduce { their results}. Also, the boost they considered in the cross section (for example, accounting for Sommerfeld enhancement), has almost exactly the same effect than lowering the mass of the WIMP particle. Furthermore, we find significant effects not accounted for { in their work} when considering the clumpiness of the DM on all its levels, but only at $z < 10$ for { a} $1TeV$ WIMP or at $z < 30$ for { a} $10GeV$ WIMP.

	\subsection{The neutral hydrogen spin temperature}
	
We now study the 21 cm signal from neutral hydrogen, and how it is modified by DM annihilation. We follow \cite{CLS} and \cite{VFMR07} in the explanation of the physics and the processes involved, and refer the reader to the above papers for further details.

The 21 cm line corresponds to the transition energy between the singlet and triplet hyperfine levels ($n=0$, $j=0$ and $n=0$, $j=1$) of the hydrogen atom. The spin temperature of the neutral hydrogen is defined as the temperature that relates the populations of singlet and triplet states:

\begin{equation}
\frac{n_{1}}{n_{0}} = 3 e^{-\frac{T_{*}}{T_{s}}},
\end{equation}

\noindent
where $T_{s}$ is the spin temperature, $T_{*}$ is the equivalent temperature of the 21 cm transition (corresponding to $T_{*} = \frac{hc}{\lambda k_{B}}$, where $\lambda = 21 cm$), $n_{1}$ is the triplet state occupation number, $n_{0}$ is the singlet state occupation number and the factor of $3$ is the degeneracy of the triplet state (i.e., 3).

In general, the spin temperature is a weighted average of the IGM kinetic temperature and the CMB temperature. In absence of additional heating mechanisms, both the IGM kinetic temperature and the spin temperature are coupled to the CMB at high redshifts ($z \gtrsim 300$), whereas at lower redshifts, the IGM cools faster than the CMB. Thus, the spin temperature will be lower than the CMB and the IGM will be visible in 21 cm absorption. 

There are two main mechanisms that couple the spin temperature and the kinetic temperature: Spin exchange collisions and Ly-alpha pumping (also called the Wouthuysen-Field (W-F) effect \citep{Hirata}). Spin exchange collisions correspond to the collisions of neutral hydrogen (responsible for the 21 cm line) with other neutral hydrogen atoms, with electrons or with protons. These collisions will tend to thermalize the populations of both levels (singlet and triplet) to the kinetic temperature of the IGM, and thus will couple $T_{s}$ with $T_{k}$. Lyman-alpha pumping, on the other hand, corresponds to the mixing (and thus thermalisation) of the singlet and triplet levels by Lyman-alpha transitions (from energy levels 1 to 2 and vice-versa) to the 2p level. Therefore, the mixing of hyperfine states through the W-F effect will depend on the intensity of the Lyman-alpha photon field, which in turn will depend on the kinetic temperature of the IGM and on the direct energy injection due to DM WIMPs. Thus, the W-F will also couple $T_{s}$ with $T_{k}$. 

At low redshift ($z<50$), the IGM density is sufficiently rarefied so as to make the spin exchange collisions too improbable, and thus $T_{s}$ will tend to follow the temperature of the CMB again. 

As the mechanisms coupling $T_{s}$ and $Tk_{IGM}$ depend on the ionization fraction $x_{e}$, and the DM energy injection rate per baryon $\dot{E}_{\chi}$, it is expected that the presence of an annihilating WIMP will alter the evolution of $T_{s}$ with redshift, and thus, the detection of this signal may help to discriminate between different WIMP scenarios.

The spin temperature at redshift z can be computed as:

\begin{equation}
T_{s} = \frac{T_{CMB}+y_{\alpha}T_{k}+y_{c}T_{k}}{1+y_{\alpha}+y_{c}},
\end{equation}

\noindent
where $y_{\alpha}$ is the W-F coupling, and $y_{c}$ is the collisional coupling to the IGM temperature. The coupling coefficients can be computed as:

\begin{equation}
y_{\alpha} = \frac{P_{10}T_{*}}{A_{10}T_{k}},
\end{equation}

\noindent
and

\begin{equation}
y_{c} = \frac{C_{10}T_{*}}{A_{10}T_{k}},
\end{equation}

\noindent
where $A_{10} = 2.85\times10^{-15}s^{-1}$ is the spontaneous emission coefficient for the transition, and $P_{10}$ and $C_{10}$ are the radiative and collisional transition coefficients respectively (they are rates, with units of $\left[s^{-1}\right]$). The collisional transition coefficient is given by:

\begin{equation}
C_{10} = k_{10}n_{HI} + n_{e}\gamma_{e},
\end{equation}

\noindent
where $n_{HI}$ and $n_{e}$ are the neutral hydrogen number density and the free electron number density respectively, and $k_{10}$ and $\gamma_{e}$ are the specific (one particle) transition coefficients for collisions with neutral hydrogen atoms and electrons respectively. We have neglected the contribution due to collisions with protons, following \cite{CLS}. The number densities $n_{e}$ and $n_{HI}$ are easily obtained in terms of the above-defined ionization fraction $x_{e}$, the hydrogen and helium abundances by number ($f_{H}$ and $f_{He}$) and the baryonic density of the universe $n_{b}$, also defined previously. We use the following formulas for the coefficients $k_{10}$ and $\gamma_{e}$:

\begin{equation}
(k_{10}/cm^{3}s^{-1}) = 3.1\times 10^{-11}\left(T_{k}/1K\right)^{0.357}exp(-32K/T_{k})
\end{equation}

\noindent
and

\begin{eqnarray}
log(\gamma_{e}/cm^{3}s^{-1}) = -9.607 + 0.5 log(T_{k}/1K) ,\nonumber\\ T_{k} \leq 1K;
\end{eqnarray}

\begin{eqnarray}
log(\gamma_{e}/cm^{3}s^{-1}) = \nonumber\\ -9.607 + 0.5 log(T_{k}/1K)exp(-(log(T_{k}/1K))^{4.5}/1800), \nonumber\\ 1K < T_{k} \leq 10^{4}K;
\end{eqnarray}

\begin{eqnarray}
log(\gamma_{e}/cm^{3}s^{-1}) =-8.102, \nonumber\\ T_{k} > 10^{4}K.
\end{eqnarray}

The fit to $k_{10}$ is given by \cite{KuhlenMadau} and the fit to $\gamma_{e}$ is given by \cite{Liszt}. 

We now calculate $P_{10}$ as:

\begin{equation}
P_{10} = \frac{16}{27}\frac{\pi J_{\alpha} \sigma_{\alpha}}{h\nu_{\alpha}},
\end{equation}

\noindent
where $J_{\alpha}$ is the isotropic specific intensity of the radiation field in the Ly-alpha line, $\sigma_{\alpha}$ is the Ly-alpha transition cross-section in the monochromatic approximation, and $\nu_{\alpha}$ is the frequency of the Ly-alpha transition photon. According to \cite{Laor}, the $\sigma_{\alpha}$ monochromatic cross-section is given by:

\begin{equation}
\sigma_{\alpha} = \frac{\pi e^{2}}{m_{e}c}f_{12}f_{se}\frac{n_{1}}{n_{1}+n_{2}},
\end{equation}

\noindent
where $f_{se}$ is the stimulated emission correction and is given by:

\begin{equation}
f_{se} = (1-e^{-\frac{h\nu_{\alpha}}{k_{B}T_{k}}}),
\end{equation}

\noindent
the first and second energy levels population ratio $\frac{n_{2}}{n_{1}}$ of the hydrogen atom is given by:

\begin{equation}
\frac{n_{2}}{n_{1}} = \frac{g_{2}}{g_{1}}e^{-\frac{h\nu_{\alpha}}{k_{B}T_{k}}},
\end{equation}

\noindent
the n-th level { degeneracy} is given by $g_{n} = 2n^{2}$, and $f_{12}$ is the oscillator strength of the transition ($f_{12} = 0.416$). Note that the factor $e^{2}$ in $\sigma_{\alpha}$ should be replaced by $\frac{e^{2}}{4\pi \epsilon_{0}}$ when using MKS units.

Finally, the specific intensity of the Ly-alpha radiation field is given by:

\begin{equation}
J_{\alpha}(z) = \frac{n_{H}^{2}hc}{4\pi H(z)}\left[x_{e}x_{p}\alpha_{2^{2}P}^{eff} + x_{e}x_{HI}\gamma_{eH} + \frac{\chi_{\alpha}\dot{E}_{\chi}(z)}{n_{H}h\nu_{\alpha}} \right],
\end{equation}

\noindent
where $x_{p} = (1 - x_{HI})$ is the proton fraction with respect to the total hydrogen number density, $\alpha_{2^{2}P}^{eff}$ is the radiative recombination coefficient to the $n=2$, $l=1$ level, given in {Table 1} of \cite{Pengelly}, $\gamma_{eH}$ is a collisional excitation rate of $HI$ atoms involving electrons, and $\chi_{\alpha} = \frac{\chi_{e}}{2}$, where $\chi_{e}$ is the excitations efficiency given by $\chi_{e} = \frac{1-x_{e}}{3}$ (here, it can be seen that $\chi_{e} + \chi_{i} + \chi_{h} = 1$, and thus, all the energy injected to the IGM by DM annihilation goes to either heating, ionization or excitation). For $\gamma_{eH}$, we use:

\begin{equation}
\gamma_{eH} = 2.2\times 10^{-8}exp\left[ -11.84/(T/10^{4}K)\right], [cm^{3}s^{-1}],
\end{equation}

\noindent
given by \cite{CLS}.

Now, we can explicitly compute the spin temperature $T_{s}$ as a function of z, and use it in the following calculations.

	\subsection{The 21 cm differential brightness temperature}

Having obtained the spin temperature $T_{s}$, we proceed to compute the 21 cm differential brightness temperature $\delta T_{b}$, defined as:

\begin{equation}
\delta T_{b} = \frac{T_{s}-T_{CMB}}{1+z}\tau(z),
\end{equation}

\noindent
where $\tau$ is the optical depth of the neutral IGM at 21(1+z) cm, and can be calculated as:

\begin{equation}
\tau(z) = \frac{3c^{3}hA_{10}}{32\pi k_{B} \nu_{0}^{2} T_{s}(z) H(z)}n_{HI}(z),
\end{equation} 

\noindent
where $\nu_{0} = 1420 MHz$ is the frequency corresponding to the 21 cm transition in the rest frame of the source. The differential brightness temperature is important because it gives a difference in brightness between the CMB and the neutral hydrogen averaged in the beam of the instrument, and will be given by the above formula when the beam is large enough that the signal equals the average global one. Thus, this is the quantity that can be measured observationally in the low frequency range, and will be used for the next section. 

	\subsection{The difference of differential brightness temperatures}

Finally, we compute the difference in differential brightness temperature, defined as:

\begin{equation}
\Delta \delta T_{b} (z) = |\delta T_{b} (z) - \delta T_{b,0} (z)|,
\end{equation}

\noindent
where $\delta T_{b,0} (z)$ corresponds to the differential brightness temperature obtained without the extra energy injection by DM annihilations. According to \cite{VFMR07}, this signal could be measured with current and future radio telescopes, such as LOFAR, 21CMA, MWA and SKA, in principle, up to the $\approx 1 mK$ level at redshifts where baryonic energy injection processes (such as PopIII star formation) are not important ($z \gtrsim 10-20$). However, { they} explain that the various foregrounds (i.e., Galactic free-free and synchrotron emission, unresolved extragalactic radio sources, free-free emission from ionizing sources, synchrotron emission from cluster radio haloes and relics) are much stronger than the cosmological signal, and are also very difficult to remove. { They} also explain some possible methods for effectively removing the foregrounds and for clearly identifying the extra energy injection signal; however, that is beyond the scope of this work and we refer the reader to their paper.

In Figure 11 and in Figure 12, we show the results we obtained for the $\Delta \delta T_{b} (z)$ curve, for the different cases we considered. In {Figure 11} we assume the maximal absorption fraction $f_{abs}=1$, and consider the different DM clustering scenarios and WIMP masses. In {Figure 12}, we consider the standard scenario for the DM clustering (adiabatic contraction with substructure), and we vary the absorption fraction, the WIMP mass, or consider the perfectly smooth case (for more details about the plots, see their captions).

\begin{figure}
\begin{center}
\includegraphics[width=0.5\textwidth]{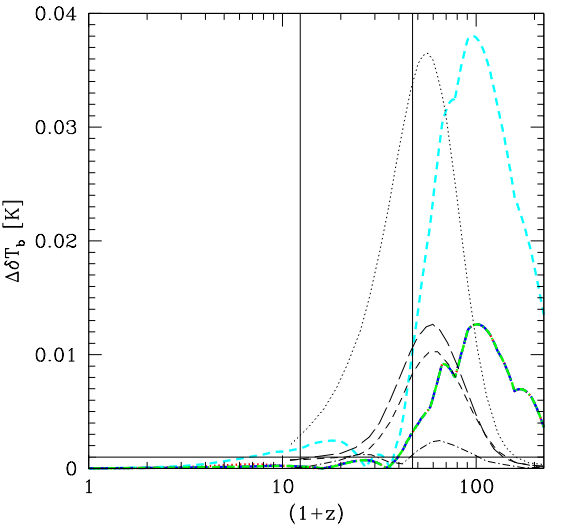}\\
\caption{{The difference of 21cm differential brightness temperature ($\Delta \delta T_{b} (z)$) for different cases, assuming a maximal absorption fraction and the clumpiness of the DM. Our results are shown as thick lines. The different line-styles are assigned in the same way as in {Figure 7}. For comparison, we also include the results obtained by Cumberbatch, Lattanzi \& Silk (2010) for the case of SUSY models with a NFW density profile, taken from the upper panel of Figure 10 of the authors, and we show them as thin lines. The line-styles for the thin curves correspond to dot-dashed for Model 4, short-dashed for Model 2, long-dashed for Model 3 and dotted for Model1. We refer the reader to Table I of the authors for a description of the different models. We also include the detectability limits here discussed, corresponding to the $1$ $mK$ threshold and to the $z = 11.35$ (PopIII stars) and $z = 46$ (minimum frequency of $30$ $MHz$) limits.}}
\label{nn}
\end{center}
\end{figure}	

\begin{figure}
\begin{center}
\includegraphics[width=0.5\textwidth]{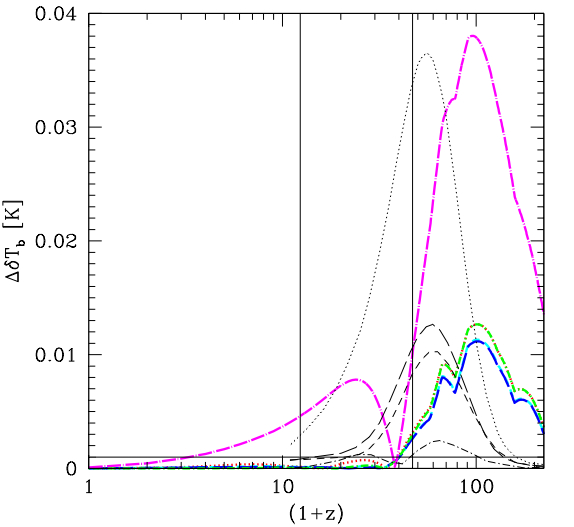}\\
\caption{{The difference of 21cm differential brightness temperature ($\Delta \delta T_{b} (z)$) for different cases, considering substructure and adiabatic contraction but with different absorption fractions, or no DM clumpiness at all. Our results are shown as thick lines. The different line-styles are assigned in the same way as in {Figure 8}. For comparison, we also include the results obtained by Cumberbatch, Lattanzi \& Silk (2010) for the case of SUSY models with a NFW density profile, taken from the upper panel of Figure 10 of the authors, and we show them as thin lines. The line-styles for the thin curves correspond to dot-dashed for Model 4, short-dashed for Model 2, long-dashed for Model 3 and dotted for Model1. We refer the reader to Table I of the authors for a description of the different models. We also include the detectability limits here discussed, corresponding to the $1$ $mK$ threshold and to the $z = 11.35$ (PopIII stars) and $z = 46$ (minimum frequency of $30$ $MHz$) limits.}}
\label{nn}
\end{center}
\end{figure}	
 
It can be seen from Figure 11, { that in the case of a $1TeV$ WIMP, all the curves corresponding to the different clustering scenarios coincide, and therefore, that} the most important factor in determining the $\Delta \delta T_{b}$ signal is the WIMP mass. For a $10GeV$ WIMP, the signal is around $\sim 40 mK$ at redshift 100, very much above the $\sim 1 mK$ sensitivity of current and future instruments quoted by \cite{VFMR07}. However, { in \cite{VFMR07}, it is also mentioned} that current and future radio frequency arrays will probably probe up to a minimum frequency of $30MHz$, thus corresponding to a 21cm emission at $z \sim 46$. Therefore, being optimistic, we consider the sensitivity window to be given by $\Delta \delta T_{b} > 1 mK$, and also $\sim 11<z<\sim 46$. Note, however, that in reality the minimum frequency, and therefore the maximum measurable redshift, would depend on the experiment (e.g., in \cite{Carilli}, the minimum frequency for SKA is given as $100MHz$, corresponding to a maximum measurable redshift of $z=13.2$, and that of 21CMA is given as $70MHz$, corresponding to $z=19.3$). { In the $1TeV$ WIMP case, the signal is within the considered sensitivity window at the $1-3mK$ level for $40 < z < 46$, corresponding to a possible detectability range of $30MHz < \nu < 35MHz$ in signal frequency. For a $10GeV$ WIMP, the signal may be detectable at the $1-3mK$ level for $11 < z < 25$, corresponding to $55MHz < \nu < 119MHz$, and at the $1-10mK$ level for $35 < z < 46$, corresponding to $30MHz < \nu < 40MHz$. However, it may be difficult to discern this signal from the backgrounds.} 

It can be seen in Figure 12, that even in the no-clumpiness case, a WIMP with mass of $10GeV$ would still be inside the sensitivity window mentioned before, due to the broad peak in the signal around $z\sim 100$ and the secondary peak around $z\sim 25$. Also, for the cases with $1TeV$ WIMP but non-maximal $f_{abs}$ or no clumpiness, the signal would be very similar to the $f_{abs}=1$ case, but with a slightly reduced amplitude, and would just barely fall in the sensitivity window (at $40 < z < 46$). 

It is important to keep in mind that, as was mentioned above, the sensitivity window here considered is only nominal, because the foreground extraction and the de-contamination of the signal may be difficult.

We also present in Figure 11 and Figure 12 the $\Delta \delta T_{b}(z)$ curves obtained by \cite{CLS} and compare them with the results that we obtain.

It can be seen from the comparison between the curves given by the authors and ours, that the dynamical range of the values for the obtained $\Delta \delta T_{b}(z)$ curves are similar, but that the particular shapes of the curves are different. This may be because the authors use an analytic approximation to $\sigma(M,Z)$, compute $f_{abs}$ self-consistently and assume a pure power-law shape for the substructure, without considering time evolution of substructures within particular haloes, and so their curves will be more soft, having less numerical noise from the integration and equation solving procedures. Also, all our curves are peaked around $z\sim100$, with a secondary peak { at $z\sim 20$} in the cases with $10GeV$ WIMPs, whereas the results from the authors are peaked around $z \sim 50-60$. { For the cases with $10GeV$ WIMPs, the secondary peak is more prominent in the case with no clumpiness at all than in the case with standard clumpiness (considering both substructure and adiabatic contraction). The reason for this is that the plotted quantity corresponds to the difference, in absolute value, between the differential brightness temperatures with and without DM annihilation, and so, in the case with no clumpiness, said difference is actually negative in the region of the secondary peak, whereas in the case that considers clumpiness, it is positive but smaller in absolute value. Thus, the prominence of the secondary peak is a feature of the more rapid cooling that occurs in the absence of clumpiness, and thus, it is the result of the DM being in haloes instead of being entirely diffuse.}  



\section{Conclusions}

In this paper, we develop formalisms for calculating the generated DM self annihilation luminosity of individual halos { and the power they receive from the radiation intensity field}, as well as the energy injection rate per baryon in the universe as a whole, accounting for the clustering of the DM at all scales but without considering Sommerfeld enhancement. We also compute the effects on the IGM and on the 21cm signal of this extra energy injection due to the DM. 

It is important to mention that we take a numeric approach for performing the calculations of the barionic clumpiness, instead of the analytic approach used by previous authors. Also, we incorporate adiabatic contraction, which had not been done so far. { In our computations, we use the prescriptions given in Section 3 for} the primordial power spectrum, the transfer function, the window function, the substructure mass function and its corresponding time evolution, the computation of the NFW parameters as a function of halo mass and redshift and the computation of the adiabatically compressed density profile for each halo, considering the critical mass criteria as given by \cite{Tegmark}.

We now give some general conclusions about the results obtained.

As discussed in Section 4.2, the luminosity generated within individual halos, considering $1TeV$ WIMPs, is not enough to self-regulate the star formation of the halo, and thus, it cannot contribute significantly to the standard feedback mechanisms (AGN feedback for haloes of mass greater than $10^{13}M_{\odot}$ and SNe feedback for haloes of mass smaller than $10^{10}M_{\odot}$).

DM clustering in general is critical in order to assess the effects of DM annihilation on the IGM and for computing the energy injection per baryon. Evidence of this are the clumpiness factors obtained in Section 4.4, which can be in excess of $10^{6}$ at redshift zero (the energy injection rate is thus boosted by this factor with respect to the case of a perfectly smooth universe). Also, in favour of the importance of clustering are the results obtained for the ionization fraction $x_{e}(z)$ and the IGM kinetic temperature $Tk_{IGM}(z)$ discussed in Section 6.1 and Section 6.2, where it is clearly seen that, for the cases assuming a perfectly homogeneous universe, and for a WIMP mass of $1TeV$, the obtained evolutions of these quantities are virtually indistinguishable from the standard case with no energy injection due to DM. For example, not accounting for { clustering}, led \cite{MFP06} to conclude that the effects of a heavy ($\sim 100GeV$) neutralino on the IGM were negligible, and it was shown by \cite{CLS} that this may not be the case by accounting for the clustering of DM. However, we found that even for our maximally clustered case (adiabatic contraction plus substructure), for a $1TeV$ WIMP, the deviations with respect to the smooth case only become significant at $z \sim 10-20$.

We find that the presence of annihilating and clumped DM may result in significant deviations in the evolutions of the inferred temperature and ionization fraction for $z \lesssim 20$. For example assuming a maximal energy absorption fraction of $f_{abs}=1$, in the scenario that takes into account the clustering of DM at all levels as well as the adiabatic contraction, a $1TeV$ WIMP with thermal relic annihilation cross-section may increase the temperature of the IGM to the $\sim 100 K$ level, or contribute to the ionization fraction rising it to a value of $\sim 5 \times 10^{-3}$ at redshift zero. { However, this effect would be hidden out by the effects of the ionizing radiation of PopIII stars for $z \lesssim 10$. At $z\sim 10$, where the effects of DM annihilation on the IGM would be measurable, the ionization fraction could be raised by a factor of 2 to $6 \times 10^{-4}$ and the temperature by a factor of 10 to $10$K. In the case of a $10GeV$ WIMP, the effects at $z\sim 10$ would be to increase the IGM temperature to $200K$ and the ionization fraction to $8 \times 10^{-3} $.} 

 We conclude that although the DM annihilation can have a significant impact on the inferred properties of the IGM, it cannot be regarded as an alternative reionization scenario, because even in the case of a $10GeV$ WIMP (with lower mass than favoured by direct and indirect detections and accelerator constraints), the value of $x_{e}$ at $z<100$ does not rise over 0.03. This is in agreement with the results obtained by \cite{MFP06}, where for all the different models of decaying or annihilating DM considered by the authors, they obtained values of $x_{e}$ at $z<100$ of at most 0.3 (for the case of axino-type Light Dark Matter with a mass of $10MeV$ and without considering PopIII stars). We note however that among the models that they considered, only the neutralino DM model is comparable to our results for the no clumpiness case. Both results are in good agreement, however we do not use exactly the same mass as they do, and we do not consider cross-section enhancement. Varying the mass of the WIMP and considering a constant enhancement in cross-section can have enormous effects on the properties of the IGM as the DM energy injection rate scales as $\propto \frac{<\sigma v>}{m_{\chi}}$ (disregarding the variation in the presence of substructure when changing $m_{\chi}$ due to free-streaming effects).

About the detectability of the $21 cm$ $\Delta \delta T_{b}$ signal, defined as the absolute value of the difference of the differential brightness temperatures in 21cm with and without DM energy injection (see Section 6.5), we can conclude considering the results that we obtain that, according to the sensitivity limits mentioned by \cite{VFMR07}, and considering a maximal absorption fraction $f_{abs}=1$, a thermal relic WIMP with mass of $1TeV$ is not likely to be detected from the global signal alone, { except perhaps at the $1-3mK$ level in the frequency range $30MHz < \nu < 35MHz$ corresponding to $40 < z < 46$} (see Section 6.5). { However, a $10GeV$} mass WIMP may be detectable { at the $1-3mK$ level in the frequency range $55MHz < \nu < 119MHz$ corresponding to $11 < z < 25$, and at the $1-10mK$ level in the frequency range $30MHz < \nu < 40MHz$ corresponding to $35 < z < 46$}, although { masses lower than $\sim 100 GeV$ are} not favoured by accelerator constraints { for the cases of KK particles and Neutralinos} (see Section 2.3). Comparing with the results obtained by \cite{CLS}, we notice that for their different SUSY neutralino models, the authors obtain $\Delta \delta T_{b}$ curves that cover a similar dynamic range than our curves. However, in the redshift range of { $11<z<46$} that is interesting for detection, they tend to predict a stronger signal than our $1TeV$ and $10GeV$ WIMP models, because their curves are centred around $z\sim 50-60$, whereas ours are centred around $z \sim 100$. This also applies to our $f_{abs}=0.01$ case, to which we should compare their results to (because the authors already incorporate an explicit $f_{abs}$ treatment and they have values $f_{abs} \sim 0.01$ in that redshift range; see their figure 12 and figure 13). Note that the authors use a lower mass $50GeV$ WIMP, but they obtain the energy and number of photons and electron-positron pairs that result from the cascades produced by the primary products of the annihilation and use these in order to compute $f_{abs}$. According to their results, all of their SUSY models would be in principle detectable, except for their model 4 (see their Table 1 for more details).

Finally, as future work, we should mention that it is in principle possible to use the DM annihilation { power} received by a halo (Section 5), together with the luminosity generated by the halo itself (Section 4.2), in order to compute the effects that the extra energy injection will have on the standard physical processes that take place in the barionic component of the halo (Intra Cluster Medium or ICM in the case of galaxy cluster-sized haloes), like for example its neutral hydrogen abundance, Bremsstrahlung luminosity, etc... Also, being able to compute the ionization fractions and temperatures of individual haloes at high redshifts, as well as their spatial clustering, may provide a way of computing the angular fluctuations of the differential brightness temperature ($\delta T_{b}$) signal for different WIMP scenarios, in order to have another test besides the predicted global signal (for example, \cite{Furlanetto} study this, but without considering substructure or adiabatic contraction).

\section{Acknowledgements}
IJA is currently supported by CONICYT (Comisi\'on Nacional de Investigaci\'on Cient\'ifica y Tecnol\'ogica - Chilean Government) and by the Fulbright commission through a joint fellowship. IJA is also supported by the Physics and Astronomy Department of the University of Southern California. Part of this work was conducted while IJA was studying at Pontificia Universidad Cat\'olica de Chile and he was supported in part by BASAL-CATA (Centro de Astronom\'ia y Tecnolog\'ias Afines - Chile) PFB-06, and Fondecyt \#1110328. Part of the calculations performed for this work were carried out using the Geryon cluster at the Centro de Astro-Ingenier\'ia of Pontificia Universidad Cat\'olica de Chile.

\end{document}